\documentclass[acmlarge,manuscript]{acmart}
\usepackage[utf8]{inputenc}

\usepackage{array,graphicx}
\usepackage{float}
\usepackage{multirow}
\usepackage[most]{tcolorbox}
\usepackage{subcaption}
\usepackage{booktabs}
\usepackage{array} 
\usepackage{amsmath}
\usepackage{booktabs}
\usepackage{multirow}
\usepackage{tabularx}
\usepackage{colortbl} 
\usepackage{geometry}
\usepackage{xcolor}

\newtcolorbox{myquote}{
  colback=gray!20,   
  colframe=black!75!white,  
  fonttitle=\bfseries,       
  left=1em, right=1em,       
  boxrule=0pt,            
  arc=8pt,                   
  auto outer arc,            
  lifted shadow={1mm}{-1mm}{2mm}{0.3}  
}

\AtBeginDocument{%
  \providecommand\BibTeX{{%
    \normalfont B\kern-0.5em{\scshape i\kern-0.25em b}\kern-0.8em\TeX}}}

\setcopyright{none}

\newcommand{\bluetext}[1]{\textcolor{black}{#1}}

\newcommand{\revision}[1]{\textcolor{black}{#1}}

\definecolor{mygreen}{RGB}{0,153,0}
\definecolor{myblue}{RGB}{14,28,156}

\newcommand \tool{PrototypeFlow}

\setcopyright{acmcopyright}
\copyrightyear{2022}
\acmYear{2022}
\acmDOI{XXXXXXX.XXXXXXX}

\acmBooktitle{The 37th Annual ACM Symposium on User Interface Software and Technology(UIST' 24)} 
\acmISBN{978-1-4503-XXXX-X/18/06}

\begin{document}


\title{Towards Human-AI Synergy in UI Design: Leveraging LLMs for UI Generation with Intent Clarification and Alignment}

\author{Mingyue Yuan}
\affiliation{%
\institution{CSIRO’s Data61 \& University of New South Wales}
  \country{Australia}
}
  
\author{Jieshan Chen}
\affiliation{%
  \institution{CSIRO’s Data61}
  \country{Australia}
}

\author{Yongquan Hu}
\affiliation{%
  \institution{University of New South Wales}
  \country{Australia}
}

\author{Sidong Feng}
\affiliation{%
  \institution{Monash University}
  \country{Australia}
}

\author{Mulong Xie}
\affiliation{%
  \institution{CSIRO’s Data61}
  \country{Australia}
}

\author{Gelareh Mohammadi}
\affiliation{%
  \institution{University of New South Wales}
  \country{Australia}
}

 \author{Zhenchang Xing}
\affiliation{%
  \institution{CSIRO’s Data61 \& Australian National University}
  \country{Australia}
}

\author{Aaron Quigley}
\affiliation{%
  \institution{CSIRO’s Data61 \& University of New South Wales}
  \country{Australia}
}

\begin{abstract}
In automated UI design generation, a key challenge is the lack of support for iterative processes, as most systems focus solely on end-to-end output. This stems from limited capabilities in interpreting design intent and a lack of transparency for refining intermediate results. To better understand these challenges, we conducted a formative study that identified concrete and actionable requirements for supporting iterative design. Guided by these findings, we propose \tool{}, a human-centered, \revision{multimodal-driven system for automated UI generation. \tool{} takes natural language descriptions and layout preferences as input to generate the high-fidelity UI design. At its core is a theme design module that clarifies implicit design intent through prompt enhancement and orchestrates sub-modules for component-level generation. 
Designers retain full control over inputs, intermediate results, and final prototypes—enabling flexible, targeted refinement by steering generation and directly editing outputs.
} 
Our experiments and user studies confirmed the effectiveness and usefulness of our proposed \tool{}.

\end{abstract}



\begin{teaserfigure}
\includegraphics[width=\textwidth]{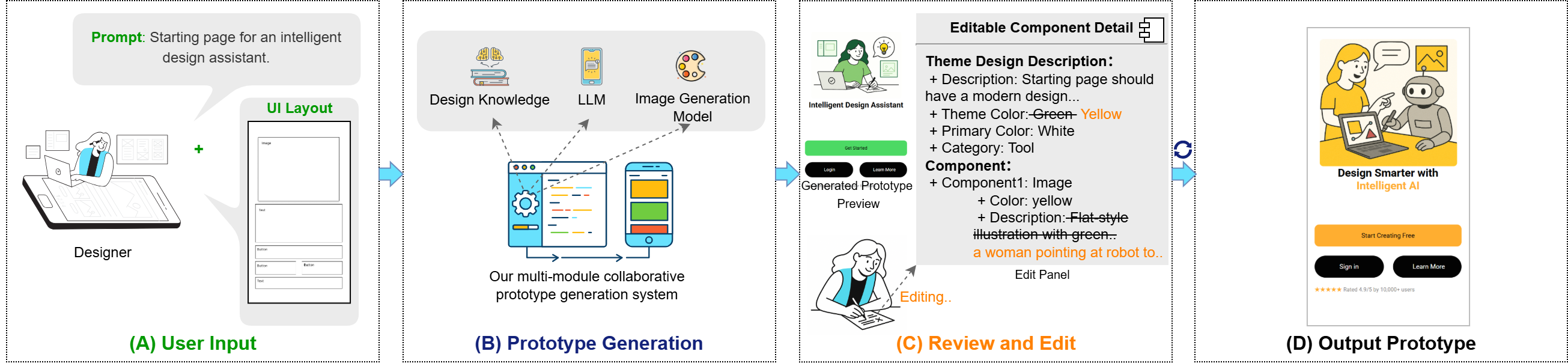}
  \caption{\revision{\textbf{Overview of the user interface (UI) prototype creation process facilitated by \tool{},} our multi-module collaborative generation system. \textcolor{mygreen}{(A)} Designers input design intentions through prompts and initial UI layouts. \textcolor{myblue}{(B)} Our multi-modal generation system produces detailed prototypes. \textcolor{orange}{(C)}  These prototypes support top-down refinement from themes to specific components,  with the option to regenerate at each level. (D) The output is a high-fidelity prototype, both visually rich and functionally grounded. }
  }
  \label{fig:teaser}
\end{teaserfigure}

\maketitle


\section{Introduction}
In recent years, the field of user interface (UI) design has seen the emergence of various tools and methods aimed at assisting designers. Notable among these are retrieval-based methods~\cite{chen2020wireframe, 2021Screen2Vec, 2021VINS, 2019Swire} and generative approaches~\cite{cheng2023play, biswas2021docsynth}. These advancements aim to streamline the design process and enhance productivity and creativity.

In designers' daily creation workflows, they typically rely on professional GUI prototyping tools, such as Sketch~\cite{sketch}, Adobe XD~\cite{adobe}, Figma~\cite{figma}. These typically offer a combination of fundamental GUI components, templates and abundant manual operations. However, they can not automatically generate customized results based on design requirements, which limits their support for the creative process and efficiency of the overall design process. 

To support the design process, many retrieval based methods have been proposed to offer inspirations by retrieving relevant UIs given designers' initial ideas, like sketch, wireframe and natural language descriptions~\cite{chen2020wireframe, 2019Swire, 2021Screen2Vec, 2021VINS, 2019Guigle, 2023Kolthoff}. 
Chen et al.~\cite{chen2020wireframe} leverages wireframes as a bridge to retrieve corresponding high-fidelity functional and visual prototypes. 
Swire~\cite{2019Swire} learns the distance between designers' sketch and high-fidelity UI design to enable sketch based search.
While Guigle~\cite{2019Guigle} use rich metadata from UI screenshots and app stores to enable natural language based query.
However, these methods often suffer from limitations in terms of visual fidelity, creativity, functional fidelity and reusability.
Recent advance in generative methods such as layout2image~\cite{biswas2021docsynth}, VAE~\cite{razavi2019generating}, MidJourney~\cite{midjourney} and stable diffusion~\cite{rombach2022high} has increased their use to support creativity, which showcase remarkable creative potential, however, their practical applications often yield unstructured and non-editable outputs.
This limits their usefulness in UI design scenarios, where further manipulation and customization are frequently needed to meet designers' specific requirements.
Additionally, small elements like icons, which carry nuanced and meaningful details, can not be handled well by these models, further exacerbating these challenges.
Moreover, both retrieved-based and generative methods often require designers to manually reconstruct these UI using professional tools to cater for their needs.

\bluetext{To address these challenges, we observe that industry tools such as Uizard~\cite{uizard}, Vercel's V0~\cite{v0dev2024}, and Figma plugins~\cite{uxpilotai} have made progress by using AI to generate initial design concepts, aiming for automated design generation. However, how AI can continuously support designers throughout the entire design iteration workflow, from concept to refinement, remains largely unexplored.
Specifically, questions remain about which aspects of the design process should remain under human control, where AI-generated processes require transparency, and how automation can be fine-tuned by humans to maximize its effectiveness throughout the workflow.}

\revision{
To gain deeper insight into human-AI collaboration challenges in AI-driven design tools, we interviewed 10 professional UI/UX designers. This study revealed five key shortcomings:
(F1) Need for streamlined design workflows supported by trend- and brand-aligned knowledge;
(F2) Need for more input control and flexible output editability in the design generation process;
(F3) Need for better support for expressing design intent through prompts;
(F4) Need for precise control in generation processes; and
(F5) Need for maintaining thematic consistency and coherence across generated components.
}

In response to these observations, we introduce \textbf{\tool{}}, a novel interactive design system that enables multi-module collaboration for UI prototype generation. 
This system empowers designers to craft high-fidelity prototypes that are editable, customisable and comprehensible at each stage of the generation process. As depicted in Fig.~\ref{fig:teaser}, \tool{} begins with two design inputs: a textual description and a wireframe layout (Fig.~\ref{fig:teaser} (A)). The text outlines the general design requirements (e.g., ``Starting page for an intelligent design assistant''), while the wireframe offers a preliminary UI layout, enabling designers to convey their initial and nuanced design concepts.

In the prototype generation phase (Fig.~\ref{fig:teaser}(B)), \tool{} employs a top-down strategy, beginning with a central module that establishes the \revision{overarching theme and performs design intent enhancement through prompt augmentation. Following the decoupling approach, the central module coordinates} the automatic, iterative creation of components, leveraging a cache pool to seamlessly integrate details from various sub-modules, each specially designed with expertise in different aspects. These specialized modules, including a text content generation module, image content generation module and retrieval-based icon module, ensure alignment with the overall design. This integrated approach allows \tool{} to maintain aesthetic consistency, elevating the quality and coherence of the final prototype.

In the review and editing phase (Fig.~\ref{fig:teaser} (C)), \tool{} offers deeper insights and detailed control within its interactive environment. \revision{When the user is dissatisfied with the theme color and the generated image, she directly changes the theme from green to yellow and revises the image description from ``Flat style illustration with green…'' to ``a woman pointing at a robot to…''. Because the theme prompt is a global parameter, this triggers regeneration of the entire prototype. } 
\revision{In comparison, for element-specific edits, the system supports localized updates. For instance, as shown in Figure~\ref{fig:webpreview}(f), if the user only updates the content field of the second row (representing the image element) in the content editor and clicks ``Regenerate Prototype'', only the image module is invoked to regenerate that particular content. }
It facilitates precise, iterative refinement, balancing automated generation with human customisation for efficient and personalized integration. Additionally, the generated prototypes (Fig.~\ref{fig:teaser} (D)) can be saved in SVG or JSON formats, highlighting the system's practicality and supporting high-quality creative output.


For evaluation, we first conducted two automated quantitative evaluations, which validate \tool{}'s ability to generate realistic and detailed UI designs.
Our ablation study further highlights the significant impact and necessity of each module.
In addition to the automated evaluations, we carried out \revision{three} user studies involving 16 participants to assess the perceived usefulness of our work. The findings suggest that \tool{} was positively received and showed significant potential in addressing the five identified challenges.
Our work includes three main contributions:

\begin{itemize}

    \item To the best of our knowledge, this work is among the first to report findings from semi-structured interviews with professional UI/UX designers working with GenAI design tools, detailing their current workflows and identifying five key shortcomings.
    \item \revision{We introduce \tool{}, a modular and interactive UI design system that enables efficient, user-centered prototype generation. The system coordinates specialized modules to regenerate the entire prototype in response to high-level changes, such as edits to the layout or theme, and also allows rapid, precise updates for component-level regeneration.}
    \item \revision{We conducted extensive experiments and user studies, which confirms the effectiveness of PrototypeFlow and surfaces actionable insights into how GenAI systems can better align with designers’ mental models and communication styles to support their creative workflow.}
\end{itemize}

\section{Related Work}


\begin{table*}[htbp]
  \centering
  \caption{Comparison of Design and Generative Tools Used in Professional UI/UX Workflows}
  \scriptsize
    \begin{tabular}{p{2.6cm}p{1.6cm}p{1.7cm}p{2cm}p{1.9cm}p{1.9cm}p{2.6cm}}
    \toprule
    \textbf{} & \multicolumn{2}{c}{\textbf{Design Tool}} & \multicolumn{4}{c}{\textbf{Generative Tool}} \tabularnewline
        \cmidrule(lr){2-3} \cmidrule(lr){4-7}
    \textbf{}   & \textbf{Adobe Illustrator} & \textbf{Figma / Adobe XD} & \textbf{Midjourney / Stable Diffusion} & \textbf{Vercel's V0} & \textbf{Uizard} & \cellcolor[gray]{0.8}\textbf{ \tool{} (Our)} \tabularnewline   
    \midrule
    \textbf{Task} & Design Tool & Design Tool & Image Generation & UI Generation & UI Generation & \cellcolor[gray]{0.8}UI Generation \tabularnewline  
    \midrule
    \textbf{Target Users} & Designers & Designers & Designers / End Users & Developers & Designers & \cellcolor[gray]{0.8}Designers \tabularnewline  
    \midrule
    \textbf{Input} & SVG & SVG & NL description, UI screenshot & NL description, UI screenshot & NL description, UI screenshot & \cellcolor[gray]{0.8}NL description, Wireframe \tabularnewline  
    \midrule
    \textbf{Output} & SVG & SVG & Image (screenshot) & Code + Rendered page & SVG & \cellcolor[gray]{0.8}SVG \tabularnewline  
    \midrule
    \textbf{Knowledge Base} & N/A & N/A & General image datasets & Confidential Knowledge Base & Confidential Knowledge Base & \cellcolor[gray]{0.8}Real-world UI, UI Components, Icon datasets, LLM-generated UI semantic datasets \tabularnewline  
    \midrule
    \textbf{\revision{Prompt Enhancement}} & N/A & N/A & Not Supported & Not Supported & Not Supported & \cellcolor[gray]{0.8}Supported \tabularnewline  
    \midrule
    \textbf{\revision{Editable Checkpoints in Generation}} & N/A & N/A &  Not Supported & Not Supported & Not Supported & \cellcolor[gray]{0.8}Supported\tabularnewline
    \midrule
    \textbf{\revision{Editable Theme Generation with Downstream Control}} &  N/A &  N/A &  Not Supported &  Not Supported &  Not Supported & \cellcolor[gray]{0.8}Supported \tabularnewline  
    \bottomrule
    \end{tabular}%
  \label{tab:Comparisontools}
\end{table*}

\subsection{GUI Design Tools and Techniques}
\label{sec:related_work_tools_compare}
In professional GUI design, tools like Sketch~\cite{sketch}, Adobe XD~\cite{adobe}, and Figma~\cite{figma} are popular due to their extensive libraries, templates, and high-fidelity prototyping features. Adobe Illustrator~\cite{adobe} is preferred for detailed illustrations. However, these tools lack robust automated generation capabilities, which limits efficiency and accessibility for designers with varying expertise. To address these constraints, research in Human-Computer Interaction (HCI) and Software Engineering (SE), along with commercial products, has focused on enhancing automation in design processes.

For creative design inspiration, 
text-based GUI retrieval, such as Guigle\cite{2019Guigle}, leverages automated crawling and natural language processing to perform efficient searches through app hierarchies. Systems like Gallery D.C.\cite{2019Gallery}, GUI2WiRe\cite{2020GUI2WiRe}, and RaWi\cite{2023Kolthoff} have further enhanced component extraction from screenshots, enabling flexible search based on dimensions, color, and text.
Visual and multi-modal retrieval methods extend this by incorporating richer inputs such as screenshots and wireframes. Methods like WAE\cite{chen2020wireframe}, VINS\cite{2021VINS}, and Swire\cite{2019Swire} bridge the gap between low and high-fidelity designs, \revision{while methods such as WireGen\cite{feng2023designing} focus on automatically generating wireframes to connect low- and mid-fidelity stages,} and Screen2Vec\cite{2021Screen2Vec} introduces multi-modal embeddings for diverse GUI content. 
In high-fidelity prototyping, tools like GUIGAN\cite{zhao2021guigan} and research by Forrest et al.\cite{huang2021creating} focus on component retrieval and arrangement for detailed prototypes. However, these retrieval-based methods still rely heavily on existing designs, limiting creative flexibility. They address early-stage inspiration but leave much of the manual effort required for design refinement and customization.

On the generative front, techniques like Layout2Image\cite{biswas2021docsynth}, VAE\cite{bengio2014auto}, MidJourney\cite{midjourney}, and Stable Diffusion\cite{rombach2022high} offer new possibilities for design generation. However, these methods often result in unstructured, non-editable outputs, making them challenging for GUI design.
Projects such as PLay\cite{cheng2023play} and DocSynth\cite{biswas2021docsynth} have made contributions to layout generation, facilitating the creation of low-fidelity prototypes. However, they provide limited support for detailed component creation or high-fidelity refinement.
ndustry tools like Uizard\cite{uizard}, Vercel's V0\cite{v0dev2024}, and Figma plugins\cite{uxpilotai} provide automated prototype generation.
\revision{Many of these tools, including Vercel and recent research systems, generate code as output~\cite{beltramelli2018pix2code, wu2024uicoder}. }
However, these approaches primarily offer initial starting points, and the missing puzzle piece is how AI can continually collaborate with designers throughout the design iteration workflow, from concept to refinement.

\revision{
While recent AI-powered design tools have made significant progress, they still present notable limitations in both input control and output editability. Most existing systems allow designers to provide either natural language prompts or image-based inputs, but rarely support multimodal input that combines wireframes with textual prompts. As a result, designers have limited ability to specify layout at the pixel level or to communicate detailed functional intent.
In response to these gaps, our system introduces multimodal \textbf{input}, enabling designers to combine wireframes and natural language prompts for precise layout specification, functional accuracy, and explicit communication of design intent.
For \textbf{output}, many previous approaches generate static images or code outputs~\cite{beltramelli2018pix2code, wu2024uicoder}, which can be difficult to refine, especially for those without programming experience. By contrast, our system produces editable SVG prototypes. This allows designers to directly manipulate visual properties and efficiently iterate, bridging the gap between generative automation and practical, high-fidelity design work. A detailed comparison of input and output modalities across leading tools is presented in Table~\ref{tab:Comparisontools}.
}


\subsection{\revision{Human-AI Interaction in GUI}}


\revision{Large Language Models (LLMs) have significantly advanced human-computer interaction, especially in graphical user interface (GUI) contexts, by enabling natural language-driven workflows and automating design processes. For instance, Wang et al.\cite{wang2023enabling}, Widget Captioning\cite{li2020widget} and Stylette\cite{kim2022stylette} showcase how LLMs facilitate intuitive conversational and command-based interactions in mobile and web UIs.} Other works, such as MenuCraft~\cite{kargaran2023menucraft}, SUGILITE~\cite{li2017sugilite}, and Duan et al.\cite{duan2024generating}, further demonstrate the versatility of LLMs in automating menu generation, task execution, and providing automated feedback on UI designs. Collaborative and educational applications are also emerging, with systems like CollabCoder\cite{gao2024collabcoder} and VIVID~\cite{choi2024vivid} highlighting LLMs’ roles in team analysis and generating dialogues from educational content.

\revision{While prior work has discussed broad HCI challenges in UI/UX, such as ethical considerations and fragmented tool ecosystems~\cite{li2024user}, as well as design workflows from ideation to mock-up~\cite{lu2022bridging}. There remains limited research on how GenAI can robustly support iterative and collaborative GUI design in professional practice.}
\bluetext{Some recent studies have begun to explore modules that clarify intents during the process of human-AI collaboration through interactions between end users and AI using natural language instructions~\cite{li2018teaching, srivastava2017joint}. SOVITE~\cite{li2020multi} expanded on this by enhancing system transparency through a mutual disambiguation pattern~\cite{oviatt1999mutual}, where inputs from one modality help clarify inputs from another, allowing for breakdown repair.}
\revision{However, interactive design generation remains under-explored and often lacks reliability and clarity. Motivated by these gaps, our work introduces a decoupled generation approach with transparent, \textbf{editable checkpoints throughout the generation process}, allowing both designers and AI to collaboratively refine and clarify intermediate results (see Table~\ref{tab:Comparisontools}).}

\subsection{\revision{Large Language Models for GUI Generation}}

\revision{Recent advances in large language models (LLMs) have opened new directions for automated GUI generation. Techniques such as chain-of-thought prompting and curriculum-driven task automation~\cite{yao2022react, gravitas2023autogpt, lu2023chameleon, wang2023voyager} allow LLMs to complete tasks with minimal human intervention. Integrating LLMs with visual models (e.g., HuggingGPT~\cite{shen2023hugginggpt}, Stable Diffusion~\cite{sd_v1_5, sd_v2_1}) further enables multi-modal workflows that bridge text and images.}

\revision{While general prompt enhancement methods~\cite{suzgun2024meta, wang2023reprompt} have improved LLM performance across a broad range of tasks, prior work on GUI generation often focuses on static prompt-based generation or one-off design variants~\cite{dang2023choice, gao2024taxonomy}, with limited attention to maintaining design coherence, enabling iterative refinement, or supporting transparent design workflows.}

\revision{Our work grounds these general advances in the context of GUI design generation by introducing a divide-and-conquer approach that translates natural language into a domain-specific language (DSL) for editable component-level specification. We then support automatic \textbf{prompt enhancement} and \textbf{editable theme generation}, enabling theme-level control and allowing designers to refine outputs for consistency in color, style, usage and layout. Finally, we employ an LLM-based controller to ensure transparency and real-time edits, thus overcoming the static and non-transparent workflows of previous approaches (see Section~\ref{sec:preliminaryStudy}, Table~\ref{tab:Comparisontools}).}

\section{Formative Study and Findings}
\label{sec:preliminaryStudy}

In this section, we conducted a study to 
\bluetext{gain insights from a professional designer’s perspective on the current design process when working with AI-powered design tools}, identify the challenges UI designers face and explore potential improvements in design tools to better support their workflows. 
\revision{
While prior studies—such as Li et al.~\cite{li2024user}, which analyzes a broad range of issues from ethical concerns to fragmented tool ecosystems, and Lu et al.~\cite{lu2022bridging}, which examines workflows related to inspiration search, mock-up generation, and iterative styling—have contributed important insights about the overall design process, there \textbf{remains a gap in understanding how GenAI tools can directly and practically support iterative design refinement in real-world professional contexts.}
Most existing research highlights the importance of iteration and encourages for improved support, but often addresses broad, end-to-end workflows or “create from scratch” scenarios without actionable and fine-grained insights. By contrast, our study focuses specifically on the \textbf{concrete, collaborative phase} of design: when objectives are already well-defined, teams must follow to established design systems or branding, and the designer’s role is to turn explicit requirements into production-ready prototypes.}

\revision{By grounding our interviews in these real-world, team-based contexts, our findings uncover some concrete challenges and needs that are not addressed in earlier, more general studies. We contribute actionable guidance on how Generative tools can better balance accuracy, creative flexibility, and seamless integration with existing workflows.}


Utilizing convenience sampling, we conducted interviews with professionals in the industry who have substantial experience in UI/UX design.
In total, we interviewed 10 UI/UX designers (2 males, 8 females) from five different companies.
These companies range from small startups to large corporations. Our participants had a diverse range of working experience: four had 1-3 years of experience, three had 3-5 years, and three had over five years in the field. Their ages varied from 23 to 34 years, averaging at 28.

The interviews were conducted through online video meetings and lasted approximately 30 minutes on average. These interviews were recorded and transcribed verbatim.
The designers were asked ten questions to assess the current status of design tools and identify areas where improvements could be made~\footnote{The questions are included in the supplementary materials}. 
Participants were questioned about their design workflow, software tools used and their benefits, the possible application of AI support at different stages, and the investigation of AI tools to streamline repetitive and less technical design tasks. 
Furthermore, we asked for their assessment of an example UI design generated by a generative model, the strengths and drawbacks of their current design tools, the demand for AI-assisted creation, and their views on aesthetics and colour requirements were also explored. 
To analyse the data, we adopted a thematic analysis approach~\cite{clarke2015thematic}. For efficiency, the coding process was carried out on a one-interview-one-coder basis. Afterwards, the two coders collaboratively performed aggregation steps. 
From our thematic analysis, we extracted 96 in-vivo codings. The study results were twofold: first, we summarized the key advantages and disadvantages of the current design and generative tools, as reported by the designers. Second, we identified five primary actionable requirements, elaborated in Section~\ref{sec:findings}.

\subsection{Findings}
\label{sec:findings}

\begin{figure}
    \centering
    \includegraphics[width=1\linewidth]{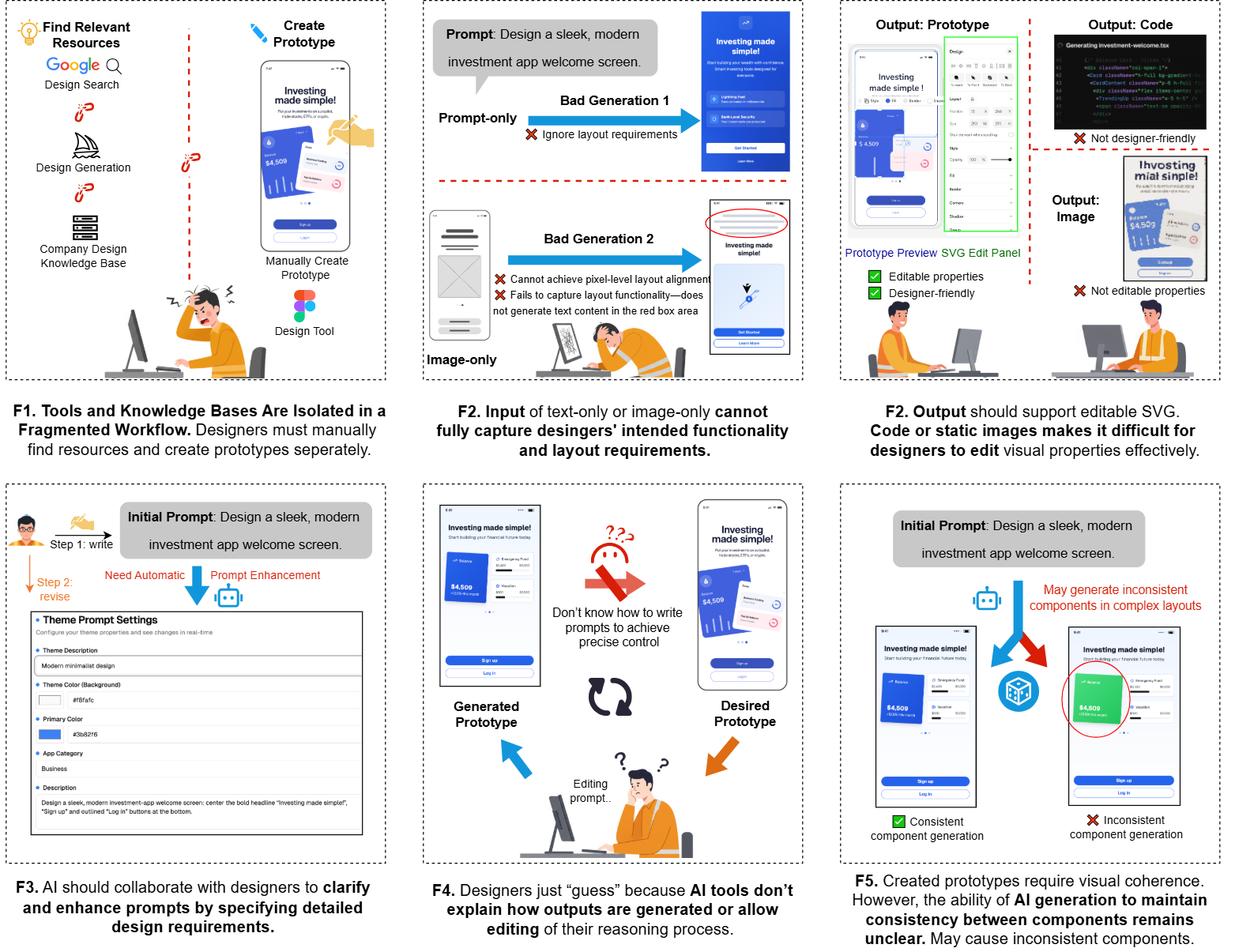}
    \caption{\revision{Illustrative Examples of Formative Study and Key Findings}}
    \label{fig:findings}
\end{figure}

Our findings indicated a prevalent use of design search tools/platforms like Google, Dribbble~\cite{dribbble}, and iconfont~\cite{iconfont}, and creative design platforms such as Figma~\cite{figma}, Adobe XD and Adobe Illustrator~\cite{adobe}. Additionally, new AI generation tools such as Midjourney~\cite{midjourney} and Stable Diffusion~\cite{sd_v1_5, sd_v2_1} have been incorporated into their daily routines. 
\bluetext{Furthermore, participants also mentioned newly released AI generative design tools, including Uizard~\cite{uizard} and Vercel's V0~\cite{v0dev2024}. However, due to their current beta status, instability, and the requirement of additional subscription fees, these tools have not yet been integrated into the designers' daily practices. }
Participants acknowledged the value of these tools in fostering creative ideas and producing well-designed UIs, while also identifying specific areas for enhancement.

Our formative interviews confirmed the limitations of current input and output modalities used by designers, highlighting the need to address design aspects that better support the generation process. Unlike commercial tools, \revision{our \tool{} focuses on prompt enhancement, enabling editable checkpoints during generation, and providing editable theme generation with downstream control.} Table~\ref{tab:Comparisontools} in Section~\ref{sec:related_work_tools_compare} compares popular design and generative tools used in participant workflows, helping to illustrate our findings and the advantages of our proposed method.
We have identified five key findings as follows:

\revision{\textbf{\textit{F1. Need for Streamlined Design Workflow Supported by Trend- and Brand-Aligned Knowledge  }}}



\revision{Participants reported a key challenge in moving from design ideas and low-fidelity wireframes to high-fidelity designs. This current process is often fragmented, requiring multiple tools and manual effort to obtain creative vision while aligning with company standards. As shown in Fig.~\ref{fig:findings} (F1), designers typically search online for implementation ideas, use tools like Midjourney for inspiration, and then consult internal guidelines to ensure consistency. This disjointed workflow—balancing trendiness, compliance, and diversity—demands considerable manual coordination. }
As one participant noted, “I regularly discover designs or elements that match what I’m looking for. However, seamlessly integrating them into my design process remains difficult. AI-powered tools should make this easier.” \revision{Participants also emphasized the need for scalable, evolving tools that can support generated UI design that reflect current trends and organizational styles. }

\begin{myquote}
\textbf{System Design Response to F1:}
\revision{Motivated by these findings, our approach directly addresses these challenges  using r \textbf{etrieval-augmented generation (RAG)}, a dynamic, training-free method that leverages the knowledge base during the generation process. This knowledge base can be continuously updated to reflect evolving design trends and company-specific requirements, enabling the generated UIs to be not only creative and visually on-trend, but also aligned with concrete design constraints. }
\end{myquote}

\textit{\textbf{\revision{F2. Need for More Input Control and Flexible Output Editability in Design Generation Process }}}


\revision{Participants expressed a shared frustration with existing AI tools, particularly regarding limited input control and output editability. They emphasized that both aspects are essential for efficiently, effectively, and accurately transforming their design requirements into deployable outcomes. }

\textbf{Input Control/Modality:}
Traditional design tools often require manual creation of designs, typically in SVG format, which is time-consuming. Emerging generative tools allow designers to input natural language descriptions or high- and mid-fidelity design images (e.g., screenshots). However, interviews with participants revealed that being able to \textit{precisely and controllably describe their design intent} is crucial. As one designer noted, \textit{``Text-only or image-only input doesn’t fully capture what I’m imagining, especially regarding the functionality of each component, which limits their practical use.'' }
\revision{For example, a text prompt such as ``Design a sleek, modern investment app welcome screen'' can lead to layouts that ignore layout requirements. Conversely, image-only input may yield visually similar designs but also fails to capture pixel-level layout or functional interactivity (see ``Bad Generation 1'' and ``Bad Generation 2'' in Fig.~\ref{fig:findings}, Input). }
Participants highlighted that combining wireframes with text descriptions allows for greater control over functionality and layout, while preserving creativity and the automation benefits of AI-powered tools.

\textbf{Output Editability/Modality:}  
\revision{Our study found that while SVG-based outputs enable property-level editing and are designer-friendly, outputs provided as code (such as those from Vercel’s V0) or as static images (e.g., from MidJourney) make it difficult for designers to refine their work—especially for those without programming experience (as illustrated in Fig.~\ref{fig:findings}, Output). }
As one participant shared,  \textit{``Getting a design from MidJourney or Vercel’s V0 is amazing, but the real work begins with the day of editing that follows.''} This highlights the need for output formats that are inherently editable and support iterative refinement.

\begin{myquote}
\textbf{System Design Response to F2:}  
\revision{Guided by these findings, our system supports multimodal \textbf{input} by allowing designers to combine wireframes with natural language prompts, enabling specification of layout at the pixel level, functional precision, and explicit communication of design intent to achieve finer-grained input control. For \textbf{output}, the system generates editable SVG prototypes, ensuring that designers can directly manipulate visual properties and efficiently iterate on their designs. These features ensure that the design process remains both accurate and adaptable, fulfilling the identified requirements for modern design interaction modalities.  }
\end{myquote}

\textit{\textbf{\revision{F3. Need for Supporting Designers in Expressing Intent Through Prompts}}}


\revision{Our study found that designers often begin with simple and high-level prompts, such as ``Design a sleek, modern investment app welcome screen.'' without specifying the detailed requirements needed for an effective design. These initial prompts typically lack information about theme colors, categories, or other visual details. Designers expressed that, due to unfamiliarity with prompt structures, uncertainty about what the AI requires, and a desire to save time when writing prompts, they frequently provide insufficient detail. Participants emphasized that AI should play an active role in helping clarify and refine these initial prompts. As shown in Fig.~\ref{fig:findings} (F3), the system could automatically generate suggested prompts for theme settings, colors, or categories, and guide designers to provide more comprehensive requirements. }

As one participant explained,  \textit{``I believe AI understands the kind of instructions it needs better than I do. If it could provide feedback, I could express my design goals more quickly and more accurately.''} This highlights the importance of collaborative intent clarification, \revision{allowing designers and AI to reach a shared understanding of design goals,} which aligns with the concept of achieving common ground in communication theory~\cite{clark1991grounding}.

\begin{myquote}
\textbf{System Design Response to F3:}
\revision{
To address this, our system includes a  \textbf{prompt enhancement} step by generating specific suggestions for theme settings, colors, and other design properties, which are features lacking in existing AI generation tools (see comparison in Table~\ref{tab:Comparisontools}, Prompt Enhancement). The system helps designers quickly specify detailed requirements, making the design process both faster and more accurate. 
}
\end{myquote}

\textit{\textbf{\revision{F4. Need for Precise Control in Generation Processes }}}

Existing studies~\cite{dang2023choice} highlight that end-to-end generation tools lack transparency and explainability, making it difficult for designers to align outputs with their true intent. 
\revision{As illustrated in Fig.~\ref{fig:findings} (F4), designers often receive prototypes that do not fully match their envisioned results. Without insight into the AI’s reasoning or step-by-step process, they are left to ``guess'' what prompt modifications will produce the desired outcome. This trial-and-error approach wastes time and reduces the effectiveness of AI-powered design. }

As summarized in Table~\ref{tab:Comparisontools} under \revision{Editable Checkpoint in Generation}, current tools do not explain how outputs are generated or allow users to edit the reasoning process. Designers in our study called for more transparent and human-centered mechanisms to clarify and refine generation steps. As one participant remarked, \textit{``If AI tools could reveal their thought processes and allow editing, it would greatly simplify tailoring the results to our needs.''}

\begin{myquote}
\textbf{System Design Response to F4:}
\revision{
To address this, our system introduces  \textbf{transparent, editable checkpoints throughout the generation process}. Designers can review and refine intermediate results instead of relying on trial and error. By adopting natural language as a domain-specific language (DSL) for global and local specifications, our approach enables collaborative, precise control over both style and components, leading to more accurate outcomes. 
}
\end{myquote}


\textit{\textbf{\revision{F5. Need for Maintaining Thematic Consistency and Coherence Across Generated Components}}}

Maintaining visual consistency, such as color, style, and layout, across all components of a prototype is critical for professional design quality, but it is often repetitive and time-consuming. 
\revision{As illustrated in Fig.~\ref{fig:findings} (F5), AI generation tools may produce inconsistent components within a project, resulting in visual incoherence. Because of AI model hallucinations or forgetting, complex or lengthy pages can contain components that deviate from the intended theme. For example, as shown in the figure, although the overall theme color is blue, a green component may be generated.}
Traditional design tools do not provide mechanisms for editable theme generation with downstream control (see Table~\ref{tab:Comparisontools}, \revision{Editable Theme Generation with Downstream Control}), which forces designers to manually adjust inconsistencies.

One participant noted, \textit{``Creating a unified look for a new app involves a lot of repetitive work.''} \revision{This highlights the need for tools that can automatically maintain design consistency and reduce the manual effort required for editing downstream components.}

\begin{myquote}
\textbf{System Design Response to F5:}
To address this challenge, our system employs an LLM-based controller for \revision{\textbf{editable theme generation with downstream control}}, ensuring consistency in color, style, and layout across all generated components. \revision{The controller transparently presents the natural language descriptions that govern design coherence, enabling designers to review, interact with, and efficiently refine consistency throughout the prototype. }
\end{myquote}

\section{Approach}
\label{sec:approach}

\subsection{Overview}


To improve the findings identified in Section~\ref{sec:preliminaryStudy} and actualize the design space outlined, we introduce \tool{}: a system that harnesses multi-module collaboration for the explainable generation of UI prototypes. This system aims to enhance \bluetext{mutual disambiguation in Human-AI Collaboration for creating high-fidelity prototypes by offering a decoupled generation process.}

While large language models excel in contextual understanding and response generation, UI generation poses unique challenges that require deep domain knowledge to ensure high-quality design. A simple end-to-end approach may fall short of capturing the complexities of UI design intent. Therefore, we adopt a divide-and-conquer, top-down strategy that decouples the generation process into sub-steps. 
\revision{This modular approach not only clarifies the generation process for both designers and the system, but also supports mutual disambiguation in Human-AI collaboration.}

\begin{figure}
  \centering
  \includegraphics[width=0.8\textwidth]{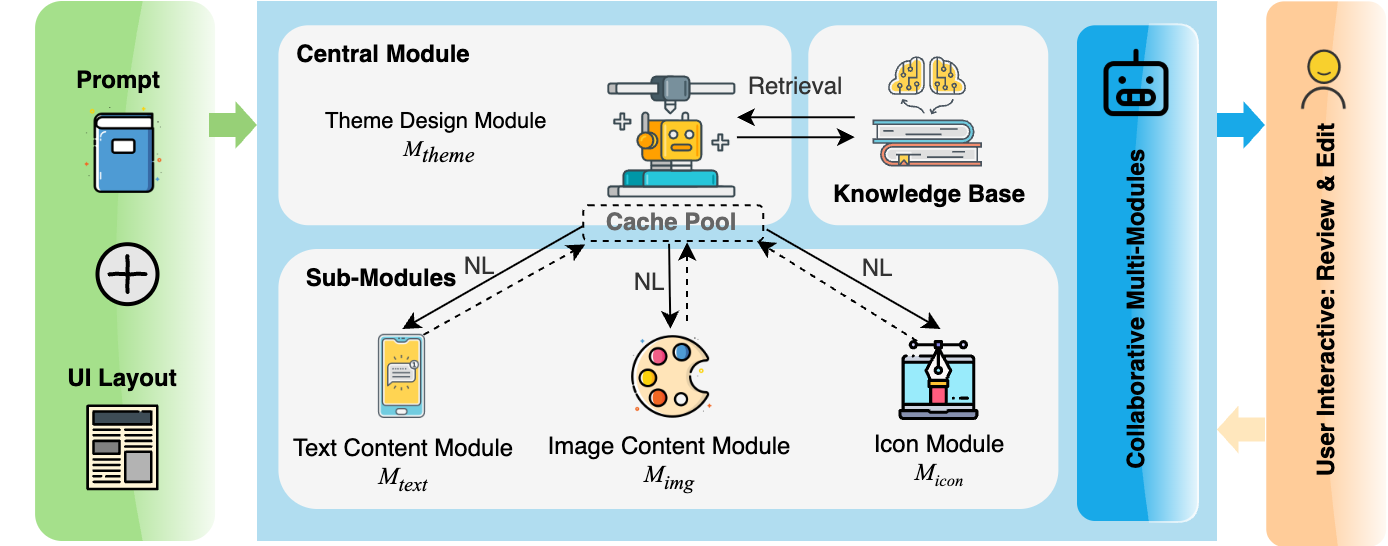}
  \caption{\textbf{Overview of \tool{} System}: This figure illustrates the main process of \tool{} in response to a designer's input and UI wireframe. \tool{} utilizes a multi-modal approach for the interactive generation of UI prototypes. It encompasses four specialized modules—Theme Design \(M_{theme}\), Textual Content \(M_{text}\), Image Content \(M_{img}\), and Icon \(M_{icon}\). The Theme Design Module \(M_{theme}\) acts as the central coordinator, steering the collaborative efforts of the three sub-modules. By leveraging a cache pool, \tool{} adeptly integrates the contributions from each module to ensure a cohesive alignment with the overall design context. This process not only generates accurate prototypes but also provides explainable intermediate results, enabling designers to conduct thorough reviews and make precise edits.}
  \label{fig:overview-MAx}
\end{figure}

\revision{As depicted in Fig.~\ref{fig:overview-MAx}, \tool{} is centrally orchestrated by the Theme Design Module \( M_{theme} \), which coordinates three specialized modules: the Text Content Module \( M_{text} \), the Image Content Module \( M_{img} \), and the Icon Module  \( M_{icon} \). The system is supported by two curated knowledge bases—one for UI layouts (22k annotated screenshots) and another for 900 diverse icons—which ground the generative process in real-world design practices.}


\revision{The theme design module serves as a central supervisor—similar to a centralized controller—that takes the user prompt and layout as input, retrieves relevant knowledge items from the knowledge base, generates a global theme description and component-level specifications, and produces a theme image to guide the image module. It ensures visual and textual consistency and coherence across the entire design, then sequentially invokes the corresponding sub-modules to generate high-fidelity content for each UI element. Finally, it assembles all generated components into the complete prototype. There is no interaction between sub-modules; all coordination and execution are managed centrally by the theme design module.}

\revision{This modular, centrally-coordinated methodology enables \tool{} to produce explainable, high-fidelity prototypes and supports iterative Human-AI collaboration. Designers benefit from both automatic generation and the ability to refine outputs, aligning results closely with their design intent.}



\subsection{Knowledge Base Construction}
\label{sec:knowledge}
Employing domain-specific knowledge can harness the creativity of Large Language Models (LLMs) while enhancing the quality of generation~\cite{wang2023voyager, lu2023chameleon}. 
We collected two kinds of knowledge, namely UI knowledge (pairs of layout/theme descriptions with local component descriptions) and Icon Knowledge (Pairs of icon SVG code with semantic descriptions) for \( M_{theme} \) and \( M_{icon} \).

\begin{table*}[htbp]
  \centering
  \caption{Blip2 model's VQA instruction templates for RICO screenshots}
  \small
    \begin{tabular}{ll}
    \toprule
    \textbf{Theme Design Attributes} & \textbf{Blip2 Instruction Templates} \\
    \midrule
    Theme Color & \textit{<Image>} \textit{``Question: What is the background color of this screenshot? Answer:''} \\
    Primary Color & \textit{<Image>} \textit{``Question: Besides the background, what's the dominant color in this image? Answer:''} \\
    Theme Description & \textit{<Image>} \textit{``Question: Can you describe this screenshot in detail? Answer:''} \\
    App Category & \textit{<Image>} \textit{``Question: Which category does this app belong to? Answer:''} \\
    \bottomrule
    \end{tabular}%
  \label{tab:vqa}%
\end{table*}%

\subsubsection{\textbf{UI Knowledge Base}} 
\label{sec:uiknowledge}
We build our knowledge base regarding the UI composition and semantic knowledge by considering two datasets (Rico~\cite{rico} and Screen2words~\cite{2021Screen2Words}) and using one large language model, Blip2~\cite{li2023blip}.
The aim of this module is to obtain two kinds of knowledge:
(1) UI Composition Knowledge: \textit{<component types><bounding boxes>};
(2) UI Semantic Knowledge: \textit{<text content/icon descriptions><high level description><theme design description>}.
An example can be seen in Fig.~\ref{fig:promptexample}(c).

\textit{\textbf{1) UI Composition Knowledge.}}
Rico Dataset~\cite{rico} is one of the most comprehensive open-source UI datasets available, which contains around 22k distinct UIs from over 9.7k Android apps across 27 categories. This dataset includes the UI screenshots and their view hierarchy information, which expose UI elements used, their attributes like text, bounds and class, and the composition of these UI elements.
We extracted class and bounds from these metadata, and formed the UI composition knowledge (\textit{<component types><bounding boxes>}) for each UI.

\textit{\textbf{2) UI Semantic Knowledge.}}
To obtain the UI semantic knowledge, we utilize the Rico dataset, Screen2Words and a visual question-answering model to obtain the fine-grained component descriptions, high-level UI descriptions and theme design descriptions, respectively.
The fine-grained component description can be obtained by parsing the text, content-description from the UI metadata from the Rico dataset. Thus, we obtained \textit{<text content/icon descriptions>}.

While Rico contains the metadata of the composition and text contents of the UI, it lacks the high-level UI description. We obtained this data through Screen2Words~\cite{2021Screen2Words}, which augments the Rico dataset by hiring crowdsourcing workers to provide ~112k high-level textual descriptions for its 22k UI screenshots. Through this dataset, we obtained \textit{<UI description>}.

Beyond UI functionality semantics, design generation necessitates thoughtful consideration of themes, colours, and the target audience. 
We adopt Blip2~\cite{li2023blip}, which is a zero-shot visual language model, to generate theme descriptions via a visual question-answering approach.
We identify four key attributes for theme design: theme colour, primary colour, theme description, and app category in Section~\ref{sec:preliminaryStudy}. 
For each attribute, we create a specific question, pairing it with a UI image, and then input it into Blip2 to derive the answer. 
The questions used to extract these attributes can be found in Table~\ref{tab:vqa}
Finally, these three descriptions are concatenated together, and form UI Semantic Knowledge: \textit{<text content/icon descriptions>< high-level description><theme design description>}.

\begin{figure}
  \centering
  \includegraphics[width=0.46\textwidth]{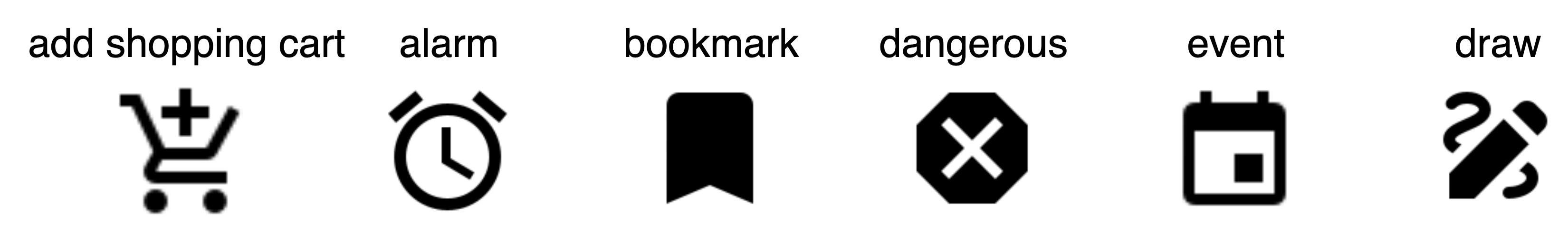}
  \caption{Examples of SVG code renderings for icons and their corresponding semantic descriptions}
  \label{fig:exampleicon}
\end{figure}

\subsubsection{\textbf{Icon Knowledge Base}}

As visual shortcuts, icons improve the user experience in UI design. They enable intuitive navigation and enhance the visual appeal. To enhance a great generation capability and enable designers to modify based on their needs, we collected an icon knowledge base in SVG format for Icon Module. 
This choice not only ensures compatibility across modern browsers but also guarantees that icons maintain their clarity when scaled.
The designer can also modify the colour, style, and shape of the icons based on their specific requirements.

We collected the data from Google Material Design Icons~\cite{materialicon}, a high-quality repository that stores over 900 diverse icons (in SVG format and with text description). These symbols adeptly convey universally recognized actions or objects, which would be ideal for design generation.
Fig.~\ref{fig:exampleicon} shows some of the collected icon examples with their corresponding semantic descriptions, such as ``add shopping cart'', ``alarm'' and ``bookmark'', to illustrate the semantic usage of these graphs.
Therefore, we obtained our Icon Knowledge Base:  \textit{<icon SVG code><semantic description>}.

\subsection{Theme Design Module}

\bluetext{The Theme Design Module functions as the central supervisor in the UI design process, leveraging domain knowledge from the UI knowledge base (Section~\ref{sec:knowledge}). Its primary role is to perform implicit intent clarification through prompt augmentation, setting the overall style and orchestrating the generation of a global theme description. By ensuring design coherence through the corresponding theme image, this dual-modality approach guarantees a high-quality UI design with consistency across all elements. The module also collaborates with designers to align and refine design descriptions, facilitating a seamless design process.}

Moreover, the Theme Design Module works in tandem with a cache pool, collaborating with other modules to refine and detail specific components of the UI.
The operational phases of the Theme Design Module comprise four key stages:
\textit{\textbf{(1) Knowledge Retrieval:}}  Accessing and utilizing domain-specific information.
\textit{\textbf{(2) Theme Description Generation:}}   
Facilitating implicit intent clarification and crafting a comprehensive and cohesive theme description.
\textit{\textbf{(3) Theme Image Generation:}}   Producing a visual representation of the theme.
\textit{\textbf{(4) Sub-module Execution:}}   Coordinating sub-modules to execute tasks, ensuring intent-design alignment throughout the generation of each component.
These stages are elaborated upon in subsequent sections, offering a detailed insight into each phase of the theme design process.



\subsubsection{\textbf{Knowledge Retrieval}}

Domain-specific knowledge enhances the accuracy of LLM-generated content. Recognizing the LLM's token input limitations and the complexities of fine-tuning, we introduce a knowledge retrieval phase, infusing external, domain-specific knowledge into our generation process.

\begin{figure*}
  \centering
  \includegraphics[width=1\textwidth]{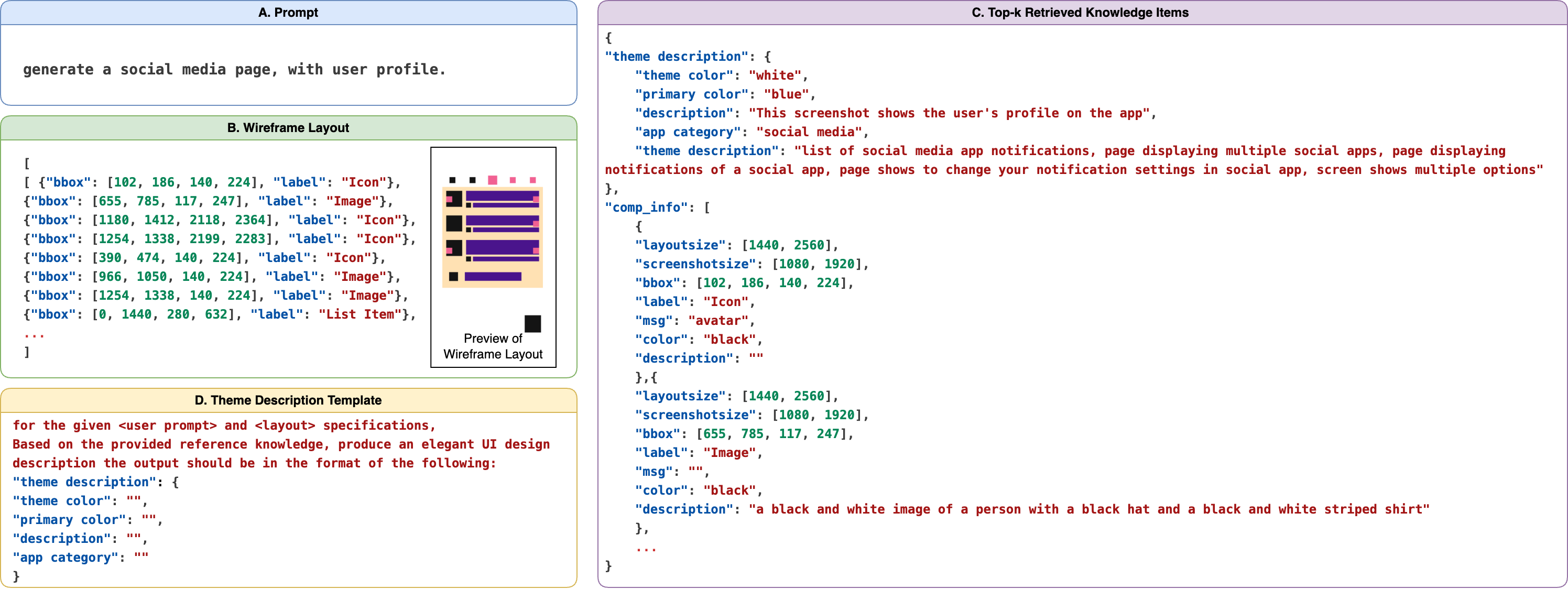}
  \caption{The prompt design for generating theme description. It consists of four parts: (A) Design Prompt, (B) Wireframe Layout, (C) Top-k Retrieved Knowledge Items, and (D) Theme Description Template. In the wireframe layout preview, different colours denote various component types as defined by Enrico~\cite{leiva2020enrico}. 
  }
  \label{fig:promptexample}
\end{figure*}

Based on the design prompt ($In_p$) and UI layout ($In_l$), which is provided in a bounding box format accompanied by component labels as an example depicted in Fig.~\ref{fig:promptexample}(B), we want to retrieve the most relevant knowledge from our large knowledge base.
To do so, we concatenate these two information together, and encode them into one latent vector $\textbf{Emb}(In)$ as the query vector, where \( In = In_p + In_l \). We use the TEXT-EMBEDDING-ADA-002 embedding model~\cite{openaiemb}).
Similarly, we also embed each piece of UI knowledge related to UI Composition and Semantic Understanding into a latent vector ($\textbf{Emb}(kb_j)$) using the same embedding model.
After that, we compute the cosine distance between the query vector and each knowledge vector, and retrieve the top-k results to instruct our multi-module system. 
We denote the retrieved knowledge as $refer_i$. 

Our preliminary experiments, alongside findings by Wang et al.~\cite{wang2023enabling}, indicate that when employing related knowledge as few-shot prompting, the initial example tends to be the most influential. Subsequent examples often provide diminishing returns in focusing the model's output. Furthermore, given the input length limitations of language models, which restrict the number of exemplars in the prompt, we set the number of references to 2 (i.e., $k=2$).

\subsubsection{\textbf{Theme Description Generation}}
\label{sec:themedescriptiongeneration}

\bluetext{The Theme Description Generation is a key part of our system, leveraging natural language descriptions as a DSL to facilitate implicit intent clarification and bridge the gap between the designer’s intent and the generated results, ensuring coherent design narratives.}

To illustrate the mechanisms driving this process, we detail a structured expression: 
Given a design prompt \( In_p \), a UI wireframe layout \( In_l \), and the top \( k \) retrieved knowledge items \( \{ \sum_{i=0}^{k} refer_i\}\) (where, \( k=2 \)), these components are concatenated with the system's theme description prompt \( P_{theme} \) to formulate a comprehensive input for our module. Formally, the amalgamated prompt \( P \) is delineated as:


\[ 
P = In_p \oplus In_l \oplus \left( \sum_{i=0}^{k-1} refer_i \right) \oplus \left( P_{theme} \right) 
\]

where \( \oplus \) symbolizes concatenation, and the summation symbol \( \sum \) here specifically indicates the sequential concatenation of retrieved knowledge items. With this integrated input at its disposal, the Theme Design Module commences the generation of the Theme Description.

Upon completing the theme description generation, the resultant theme description is denoted by \( Res_{theme} \). 
We show an example in Fig.~\ref{fig:promptexample}, where the prompt \( In_p \) is depicted in part (A), the wireframe layout \( In_l \) in part (B), the top \( k \) retrieved knowledge items \( \{ \sum_{i=0}^{k} refer_i\}\) in part (C), and the system's theme description prompt \( P_{theme} \) is presented in part (D).

\subsubsection{\textbf{Theme Image Generation}}
\label{sec:themeimagegeneration}
The objective of generating a theme image is to visually guide the overall design and ensure intent alignment. allowing the sub-module to generate coherent and consistent design.

We consider the Stable Diffusion model~\cite{rombach2022stablediffusion}, a state-of-the-art text-to-image generation, as our main model.
The main assumption of the stable diffusion model is that given a random image, we can gradually denoise it to ultimately obtain the meaningful UI design that we want. 
Therefore, the primary objective of stable diffusion is to predict and progressively eliminate noise from the initial image. 
However, as this model only considers the text condition, it suffers from limited control over the spatial composition of the image, a crucial aspect of our UI design generation. 
To address this, we integrate ControlNet~\cite{controlnet}, which augments the diffusion model by providing enhanced spatial control over each module.
In detail, the stable diffusion model contains three parts: Text Encoder, Denoising Module (i.e., UNet) and Autoencoder Decoder. 

\begin{figure}[h]
  \centering
  \includegraphics[width=0.7\textwidth]{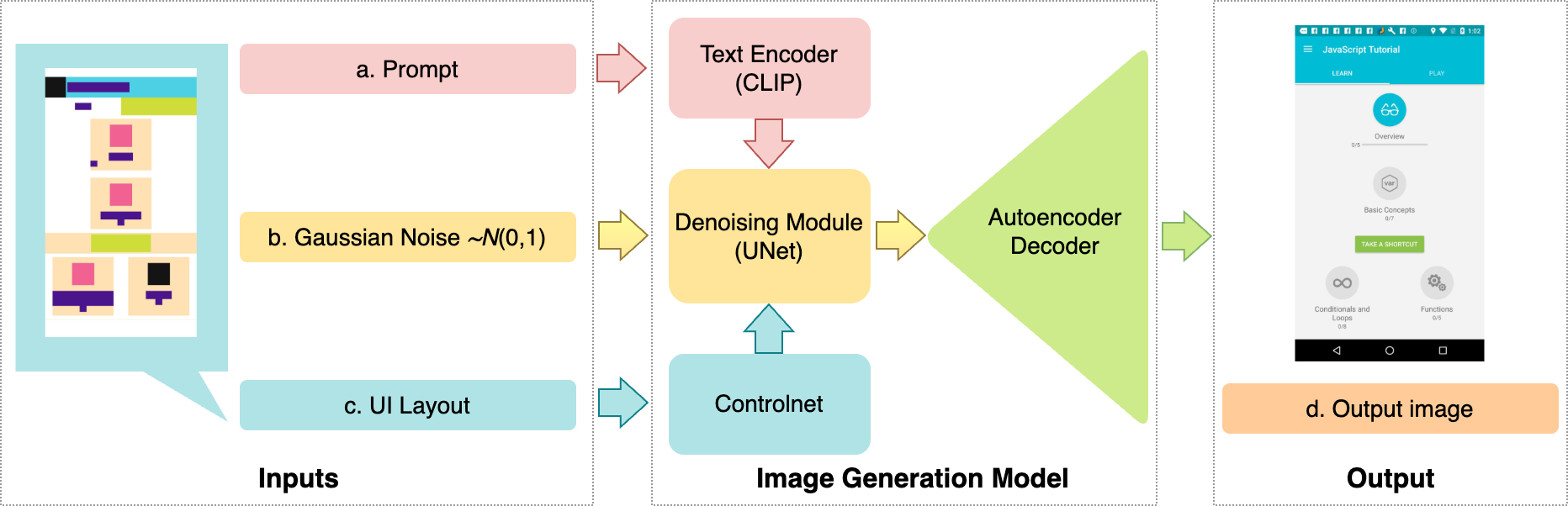}
  \caption{Layout-guided text-to-image Generation. 
  }
  \label{fig:imagegeneration}
\end{figure}

The Text Encoder encodes the prompt, Denoising Module denoises the original random images in several rounds, and the Autoencoder Decoder finally generates the image.
ControlNet controls the generation by manipulating the denoising module by inserting an additional spatial condition to UNet. 
We employ UI layout as the spatial condition. 

Specifically, as seen in Fig.~\ref{fig:imagegeneration}, this theme image generation module takes three inputs: (a) prompt (the text condition);(b) a latent image generated from Gaussian noise $\sim N(0,1)$ (the original image), and (c) UI layout (the spatial condition).
We first put \textit{(a) prompt} into the text encoder of diffusion model, pre-trained CLIP ViT-L/14~\cite{frans2022clipdraw} text encoder, and feed \textit{(c) UI layout} to ControlNet, to obtain their embeddings respectively. We then gradually denoise \textit{(b) the latent image} through the Denoising Module with the control from ControlNet. Finally, we obtain the output by feeding the denoised image into the Autoencoder Decoder.

In addition, as the stable diffusion model is originally trained on LAION-5B,~\cite{schuhmann2022laion} and faces obstacles when generating UI images, which requires a different domain knowledge from general images~\cite{chenObjectDetection}, We finetuned the model using the datasets of UI screenshots and their complementary high-level descriptions collected in Section~\ref{sec:knowledge}.

\subsubsection{\textbf{Sub-module Execution}}

During the sub-module execution phase, the Theme Design Module identifies the optimal sub-module corresponding to the component type. 
We considered 13 component types, as detailed by the Rico dataset. To elaborate, the Text Module handles ``Text Button'' and ``Text'' components. The Image Module is entrusted with ``Image'' and ``Background Image'' components, and the Icon Module focuses on ``Icon'' components. 
For other component types, we seek to render them editable, drawing insights from RaWi \cite{kolthoff2023data}. 
The colour for any component is determined by identifying the dominant RGB colour from the image region's histogram and then representing it in HTML code.

The dynamic between the central module and the sub-modules is essential to our system's functionality. When a central module engages a sub-module, it uses a combination of past results and a cache pool to inform the sub-modules prompts:

\begin{equation}
Cache_t = Res_{t-1} + Cache_{t-1}
\end{equation}

\begin{equation}
p_{t+1} = p_{sub} + Cache_t
\label{eq:subprompt}
\end{equation}

In this context, \(Res_{t-1}\) is the output from the sub-module for the \(t-1^{th}\) component. 
The \(Cache_t\) represents a cache pool that integrates the previous result with accumulated knowledge from earlier iterations. This cache pool serves as an essential memory function, retaining the design context and facilitating multi-round interactions by allowing the system to ``remember'' past interactions and decisions. 
Meanwhile, \(p_{t+1}\) functions as the prompt for the \(t-1^{th}\) component's sub-module interaction, incorporating both the specific prompt \(p_{sub}\) for the current sub-module and the cumulative knowledge in \(Cache_t\). This mechanism ensures that each sub-module's action is informed by the historical context, enabling consistent design, \bluetext{intent alignment} and coherent multi-conversation interaction. \revision{The caching mechanism is maintained and accessible by the Central Module only, who is responsible for triggering the corresponding module by providing relevant information and receiving data from each module and storing it in the cache pool.}

Further details on how these prompts are formulated for specific sub-modules, such as the Text Content Module and Icon Content Module, are outlined in Sections \ref{textagent} and \ref{iconagent}, respectively. 

\subsection{Text Content Module}
\label{textagent}
The primary role of Text Content Module is to generate textual information tailored to specific GUI components. 
We use GPT-4~\cite{gpt4}.
The system prompt for this module, represented as \(p_{text}\), is: \textit{``Based on the theme description and relevant details, provide a text content recommendation for the designated position at [bbox].''} 
In alignment with equation \ref{eq:subprompt}, the execution prompt of the Text Content Module obtains its value from the central module's cache, denoted as \(Cache_{t-1}\), and is subsequently concatenated with \(p_{text}\), to ensure consistency in the system.
This systematic approach ensures that the system consistently produces text content that seamlessly integrates with the overall theme of the GUI component.

\subsection{Image Content Module}
\label{imgagent}
To enhance the generation quality of local image-associated components and maintain the consistency of the generated outcomes, we also deploy our adaptively fine-tuned stable diffusion model in the Image Content Module \( M_{img} \). 
We reuse the Stable Diffusion model finetuned in Section~\ref{sec:themeimagegeneration} but disable the ControlNet module.
Rather than using the latent image generated from Gaussian noise, we extract the area of the image component from the theme image as the input (b).
In addition, we use the image description from the generated theme description as the prompt (a).
By harmonizing both textual and visual signals, we can guarantee that the produced content aligns seamlessly with the primary theme design intent.

\subsection{Icon Module}
\label{iconagent}
The Icon Module is crucial for selecting appropriate icons and integrating them into the graphical user interface components.
In addition to acting as intuitive visual cues, well-designed icons can improve comprehension and the overall user experience. The system prompt for the Icon Module, \(p_{icon}\), is: \textit{``In reference to relevant information and taking into account its positioning at [bbox], and based on the theme description, propose an indicative phrase like ``msg'' for the ``Icon''.} 
As shown in equation \ref{eq:subprompt}, the Icon Module's execution prompt is based on the central module's cache, \(Cache_{t-1}\), combined with \(p_{icon}\). This approach ensures the icons selected match the GUI design semantically and visually. The Icon Module then retrieves the optimal icon SVG code from the knowledge base, corresponding to the generated semantic phrase.

\begin{figure*}[!h]
  \centering
  \includegraphics[width=0.94\textwidth]{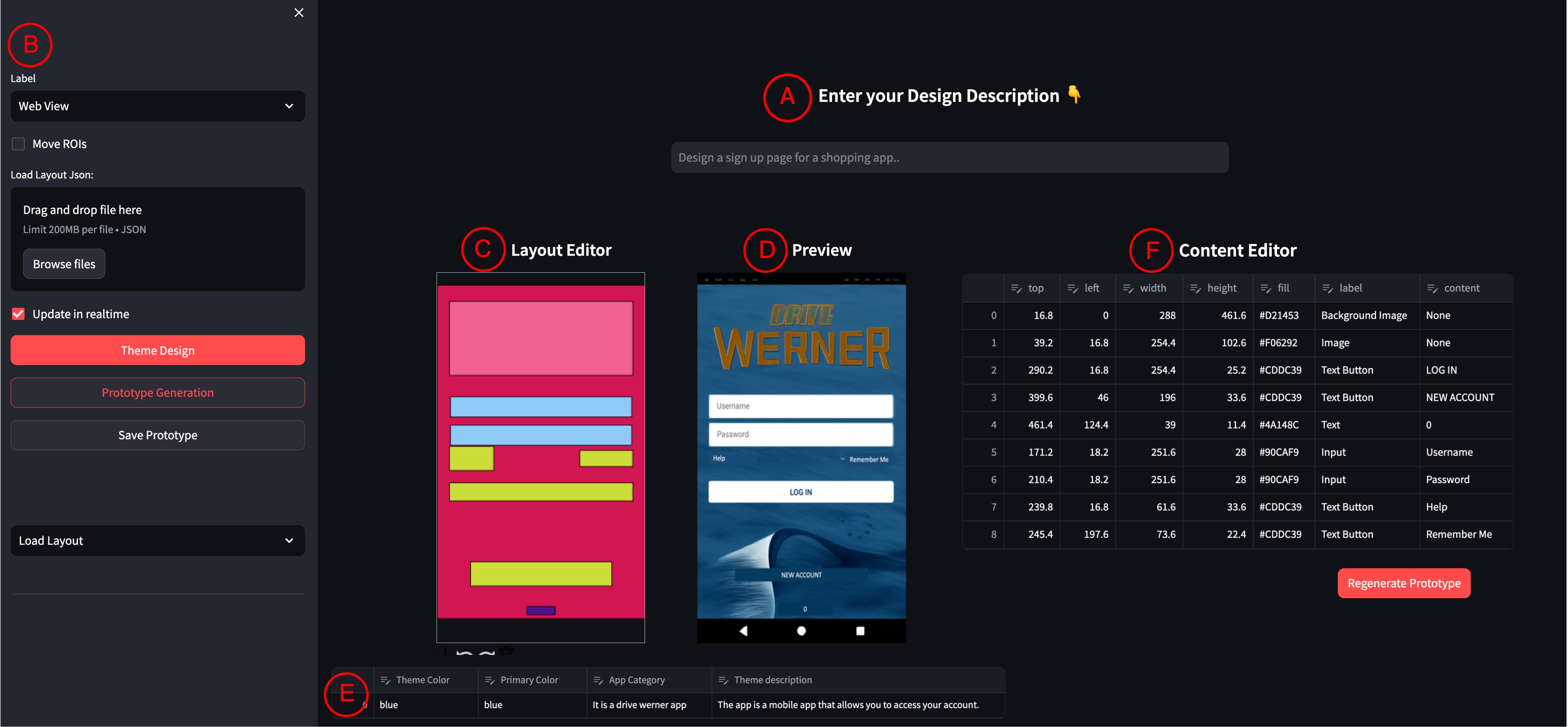}
  \caption{Interactive Interface of the \tool{} Online Prototyping Editor}
  \label{fig:webpreview}
\end{figure*}

\section{Implementation}

In the implementation of the interface, we developed a web-based rapid prototyping editor using HTML5 and JavaScript, to offer a live preview of components and an interactive editing environment. Motivated by the findings from Section~\ref{sec:preliminaryStudy}, \tool{} is adept at processing prompts and wireframe inputs to respond to \textbf{F2}. 
The Online Editor is designed for easy input, with an editable text box that let designers create task prompts seamlessly.

Adding and configuring UI elements is made simple: designers select an element type and can then resize or reposition it using a convenient drag-and-drop interface, as shown in Fig.~\ref{fig:webpreview}. The editor is structured into distinct parts for optimal user experience: Part A allows for the input of design descriptions, Part B serves as the user selection and edit panel where components such as ``Web View'' and ``Button'' can be chosen, and layout actions like ``Create Layout'' or ``Load Layout'' are available. Manual adjustments to layout size and positioning are done in Section C, while Section D provides a real-time preview of both the layout and the generated results.

Sequentially, in response to \textbf{F3} and \textbf{F5}, activating the "Theme Design" function, the generated theme descriptions are displayed in Section E, and theme image are displayed in Section D. This enhances transparency and explainability in the Human-AI collaboration process, enabling mutual disambiguation. Hitting the ``Prototype Generation'' button makes our system produce a detailed prototype in Section D, with the components' explainable contents displayed in Section F, aligned with \textbf{F4}. Designers can then save their prototype with the ``Save Prototype'' button, exporting it as a JSON file for further use.

Beyond Section D's preview feature, every part of our system is designed to be editable, empowering designers to efficiently refine and produce their final prototype. Upon generating components via our multi-module system, designers can refine both the intermediate and final outputs. The system allows for real-time visualization of the application design, and any changes made by the designer are instantly updated in the system with the ``Update Edit Result'' button, which simplifies group text modifications and streamlines the design refinement process.

For the LLM model's deployment, we utilized OpenAI's gpt-4~\cite{gpt4} APIs for textual generation within our multi-module system and paired it with the text-embedding-ada-002~\cite{openaiemb} API for text embedding. All temperature parameters were set to 0.

Regarding the image generation model, we used the runwayml stable-diffusion-1-5~\cite{sd_v1_5} checkpoint from HuggingFace to finetune the model, and the Rico dataset.
The dataset was divided into training, validation, and testing sets, consisting of 15,743 UIs, 2,364 UIs, and 4,310 UIs, respectively, following Screen2Words~\cite{2021Screen2Words}.
Training images were adjusted to $512 \times 512$ resolutions, and optimization was achieved using the AdamW algorithm with a learning rate set at 1e-5 and a batch size of 1. This training utilized an Nvidia 3090 GPU 24GB VRAM.

\section{Evaluation}

In this section, we evaluate the performance of generated prototypes.
We consider two research questions (RQs):
\begin{itemize}
    \item \textbf{RQ1}: How does our approach perform against existing models in terms of quality and diversity of the generated UI design?
    \item \textbf{RQ2}: How do \tool{}'s individual modules influence its performance in quality and diversity? 
\end{itemize}

\subsection{Evaluation Data}

From Section~\ref{sec:uiknowledge}, in total, we collected a set of 3,738 UI textual descriptions, their corresponding wireframes and high-fidelity UI design screenshots. 
For validation purposes, the corresponding UI screenshots are treated as the ground truth, and we experimented on the test set. 
Images are resized to dimensions of $512 \times 512$.

\subsection{Metrics}

To assess the quality and diversity of generated UI design, we utilize two metrics: Fréchet Inception Distance (FID)~\cite{2017GANs} and Generation Diversity (GD)~\cite{2019AIsketcher}, which are commonly used in the image generation task~\cite{2019AIsketcher, 2020DoodlerGAN}.

\textbf{Fréchet Inception Distance (FID)}~\cite{2017GANs} serves to quantify \textit{how closely the generated images resemble real ones}. This metric computes the statistical difference between distributions of generated images and their real counterparts.
Specifically, the FID score is defined as:

\begin{equation}
FID = ||\mu_r - \mu_g||_2^2 + Tr(\Sigma_r + \Sigma_g - 2(\Sigma_r \Sigma_g)^{1/2})
\end{equation}

Here, $\mu_r$ and $\mu_g$ denote the mean values of the 2048-dimensional activations of the Inception-v3 pool3 layer for real ($r$) and generated samples ($g$), respectively. Meanwhile, $\Sigma_r$ and $\Sigma_g$ represent their respective covariances. 
A \textit{lower} FID score suggests that the generated images' distribution \textit{more closely matches} that of the real images, indicating superior quality and diversity. 
To compute the FID score, an equal number of real and generated images are fed into the Inception-v3~\cite{inception} network. \revision{This standard evaluation protocol allows for consistent and comparable results within the image generation community.}

\textbf{Generation Diversity (GD)}~\cite{2019AIsketcher} 
\revision{measures the low-level visual diversity among generated prototypes, rather than high-level semantic differences. This metric helps ensure that outputs are not overly uniform or lacking in content. GD is particularly valuable for detecting failure cases where the generative model produces nearly blank images or outputs with minimal color variation (e.g., completely black or white images, which result in low GD values). }
It calculates the pairwise distances between different UI designs within a generated set. Utilizing Perceptual Hashing~\cite{Zauner2010PerceptualHash}, the metric computes these distances, with larger average distances indicating a broader variety of designs within the generated set. The formula of GD is:
\begin{equation}
GD = \frac{1}{N(N-1)} \sum_{i=1}^{N} \sum_{j=1, j\neq i}^{N} d(F_i, F_j)
\end{equation}
where \(N\) is the total number of UI designs in the generated set, \(F_i\) and \(F_j\) represent the feature vectors of the \(i\)-th and \(j\)-th UI design, respectively, and \(d(\cdot)\) denotes the Euclidean distance. 
\textit{A higher GD means the generated images are more diverse}.
For a consistent quantitative analysis, we use the same Inception-v3 model to extract features for both FID and GD evaluations.

\begin{table*}[t]
\begin{minipage}{0.45\textwidth}
\centering
\caption{Comparative analysis of FID and GD scores among various models with and without the ControlNet. A lower/higher FID/GD means the generated images are more realistic/diverse.}
\label{tab:RQ1}
\begin{tabular}{lcc}
\toprule
Model & FID↓ & GD↑\\
\midrule
stable-diffusion-1-5 (w/o ControlNet) & 69.48 & 15.93 \\
stable-diffusion-2-1 (w/o ControlNet) & 67.15 & 15.42 \\
\textbf{\tool{} (w/o ControlNet)} & \textbf{33.08} & \textbf{15.95} \\
\midrule
\\
stable-diffusion-1-5 (with ControlNet) & 54.42 & 11.48 \\
stable-diffusion-2-1 (with ControlNet) & 57.23 & 11.14 \\
\textbf{\tool{}} & \textbf{23.76} & \textbf{13.98} \\
\bottomrule
\end{tabular}
\end{minipage}%
 \hfill 
\begin{minipage}{0.45\textwidth}
\centering
\caption{Results of the ablation study for the different modules in \tool{}.}
\label{tab:RQ2}
\begin{tabular}{lcc}
\toprule
Model & FID↓ & GD↑\\
\midrule
\tool{}  & \textbf{23.76} & \textbf{13.98} \\
\midrule
-Retrieved Knowledge Items  & 42.56 & 12.14 \\
-Theme Description Generation  & 28.43 & 11.77 \\
-Theme Image Generation  & 33.08 & 12.95 \\
-Text Content Module  & 24.06 & 13.78 \\
-Image Content Module  & 24.71 & 13.38 \\
-Icon Content Module  & 24.32 & 13.41 \\
\bottomrule
\end{tabular}
\end{minipage}
\end{table*}

\subsection{RQ1: Comparisons to existing image generation models}

\subsubsection{Baselines}

We consider two state-of-the-art image generation models: \textbf{stable-diffusion-1-5}~\cite{sd_v1_5} and \textbf{stable-diffusion-2-1}~\cite{sd_v2_1}. Stable-diffusion-1-5, released in October 2022, is a widely accepted and stable version of the model. The subsequent version, 2-1, was unveiled in December 2022. It enhances the generation of images with greater diversity and realism, particularly for people, designs, and wildlife. Furthermore, it offers support for non-standard resolution generation.
As these two baselines do not incorporate ControlNet module, we consider variants of both by integrating ControlNet, denoted as \textbf{stable-diffusion-1-5 (with ControlNet)}, \textbf{stable-diffusion-2-1 (with ControlNet)}. 
In addition, we also employ an ablated version of our approach, \textbf{\tool{} (w/o ControlNet)} as a baseline.

\subsubsection{Results}

As seen in Table \ref{tab:RQ1}, several noteworthy observations can be made:

\textit{\textbf{Dominance in Quality.}} Our model, \tool{}, consistently outperforms both baseline models in terms of FID scores, regardless of whether ControlNet is utilized. Specifically, a lower FID score suggests that the distribution of generated images more closely matches that of real images. This indicates that \tool{}'s outputs are more realistic, evident from its significantly reduced FID scores: 33.08 without ControlNet and an even lower 23.76 with ControlNet.

\textit{\textbf{Consistent Diversity with Details.}} GD scores reflect the model's ability to produce varied yet detailed UI designs. A higher GD indicates more detail, suggesting that the designs are diverse and intricate in their presentation. Without ControlNet, \tool{} achieves a GD of 15.95, showcasing a balanced performance between quality and detailed variety. When integrated with ControlNet, the model achieves a commendable GD of 13.98. Although slightly reduced, this score underscores \tool{}'s ability to produce diverse and detailed outputs, even within layout constraints. Notably, \tool{}'s GD scores, both with and without ControlNet, surpass those of the baseline models. This increase in GD values emphasizes our model's superior capability in producing designs that are varied and enriched with details compared to its peers.

\textit{\textbf{Impact of ControlNet.}} Incorporating ControlNet results in noticeable improvements in FID scores for all models. For \tool{}, the FID shows an enhancement of 10.68, representing a substantial 32\% improvement. However, the slight decrease in GD, from 15.95 to 13.98, suggests that while ControlNet enhances image realism, it might limit the detail in generative diversity. This trade-off between quality and diversity is anticipated since layout constraints naturally reduce the range of potential outputs.

\textit{\textbf{Comparing Our Fine-tuned Model with Baselines.}} Contrasted with the baseline models, the advantages of our fine-tuned model become clear. In both scenarios, with and without ControlNet, \tool{} achieves superior FID scores, highlighting its excellence in UI design generation. The consistent GD scores, even surpassing some baselines, confirm the model's capacity to generate designs that are of high quality and rich in detail.
We will provide qualitative evaluation through a user study in Section~\ref{sec:userstudy1}.

\begin{myquote}
\textbf{Answer to RQ1:}
Our \tool{} outperforms the baseline models in terms of both quality and diversity.
The addition of ControlNet optimizes this performance further by introducing layout constraints, reinforcing its potential for generating realistic and detail-oriented UI designs.
\end{myquote}

\subsection{RQ2: Ablation Study}
\label{sec:rq2}

\subsubsection{Baselines and Ablation Strategy}

To better understand the interaction and individual impact of the decoupled generation mechanisms of \tool{}, we perform an ablation study, sequentially removing each module and evaluating the effect.
As seen in Table \ref{tab:RQ2}, we carefully crafted six ablations. 
These modules span across the four systematic phases, namely \textit{Knowledge Retrieval}, \textit{Theme Description Generation}, \textit{Theme Image Generation}, and \textit{Sub-module Execution}. The \textit{Sub-module Execution} phase, being more granular, was further subdivided to examine the effects of the \textit{Text Content Module}, \textit{Image Content Module}, and \textit{Icon Content Module} individually.

\subsubsection{Results}

By studying the performance impact when a specific module is absent, we can measure its individual contribution. For instance, removing the \textit{Knowledge Retrieval} module lets us understand the contribution of our knowledge base in shaping the generated UI designs.
Furthermore, these baselines serve as a means to pinpoint the robustness of our model. A model that exhibits minimal degradation in performance across different scenarios showcases its robustness and flexibility.

\begin{figure*}
  \centering
  \includegraphics[width=0.90\linewidth]{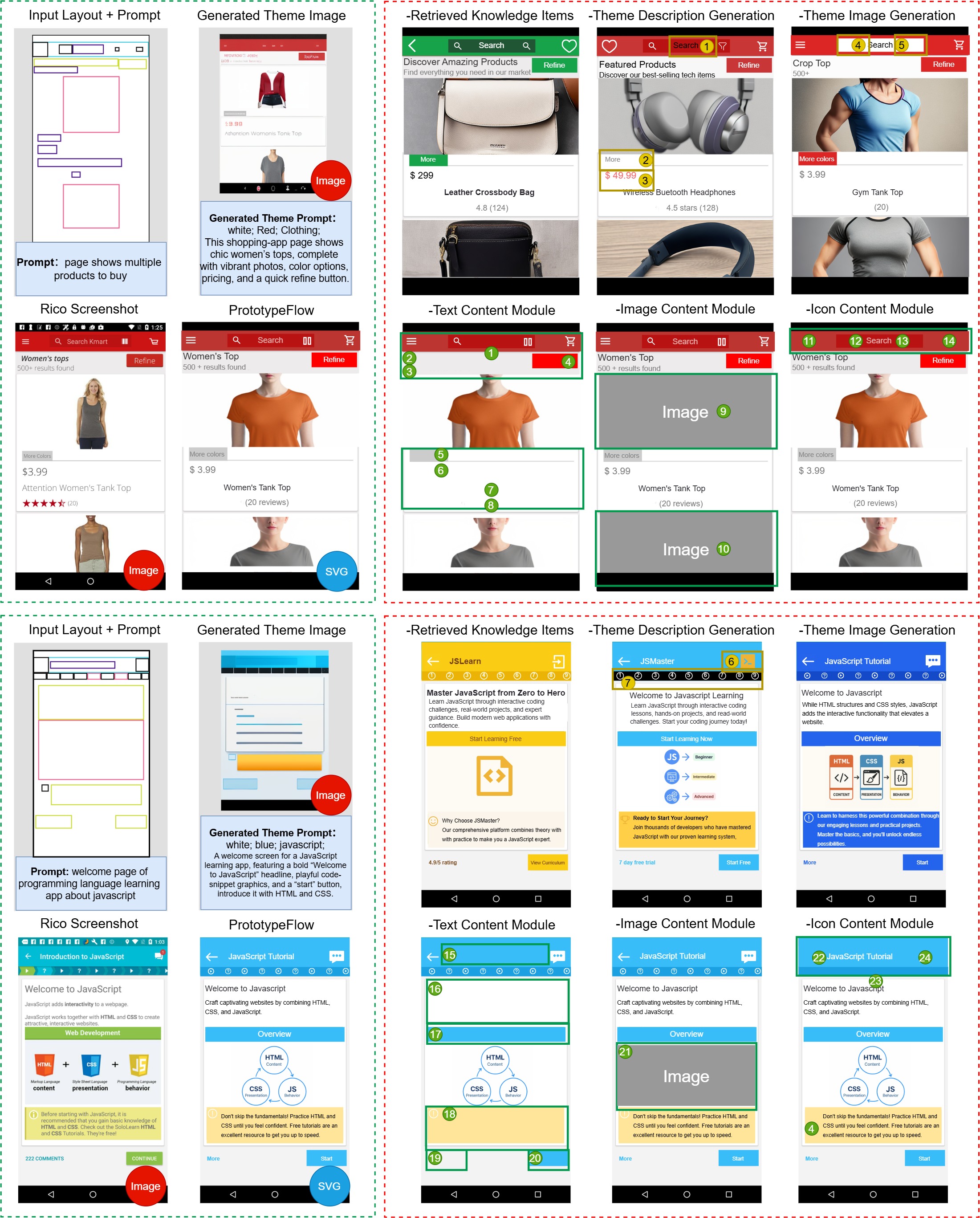}
    \caption{\revision{Ablation Study: GUI examples generated from input layout and prompt, with corresponding Rico screenshots, outputs from our \tool{}, and results from ablated modules (Retrieved Knowledge Items, Theme/Content Modules). Green boxes and numbered circles indicate missing components; yellow indicate components with theme color inconsistencies, both due to module removal.}}
  \label{fig:rq2-ablation}
\end{figure*}

\textit{\textbf{Knowledge Retrieval.}} Fundamental to \tool{}, this module imports essential data from our knowledge base. Excluding it results in an increase of 18.8 in the FID score (from 23.76 to 42.56). This represents the most significant decline in generation quality, underscoring that our knowledge base is pivotal in laying the foundation for high-quality design outputs. 

\revision{Fig.~\ref{fig:rq2-ablation} shows that removing \emph{Retrieved Knowledge Items} increases randomness in theme color and component selection. This increased diversity is highly valuable during early ideation and when creating designs from scratch, as it provides designers with a broad range of options. However, it conflicts with designers’ expectations once themes and functional requirements have been established, because the output no longer reflects the existing design knowledge base. In a real production scenario, this would force designers to perform additional manual editing.}

\textit{\textbf{Theme Description Generation.}} At the heart of generating nuanced and relatable user interfaces, this module crafts detailed textual narratives that align seamlessly with the intended theme. The act of removing it results in an FID score of 28.43 and a GD of 11.77. The degradation of FID and GD scores distinctly shows that the module plays an indispensable role in maintaining design excellence. The Theme Description Generation acts as a directive force, ensuring that the generated outputs are not only contextually meaningful but also visually relevant. Without this narrative guide, the designs might lack depth and context, affecting the overall user experience.

\revision{When \emph{Theme Description Generation} is removed in Fig.~\ref{fig:rq2-ablation}, the overall color palette remains correct because the \emph{Theme Image} still enforces the primary theme color (red in the first example, blue in the second). However, this ablation introduces localized color disharmonies in specific components. Yellow boxes 1–3 and 6–7 highlight elements whose colors no longer align with the top bar or theme: for instance, black text appears where white is expected for annotated (1), a yellow icon instead of white for annotated (6), and a black icon background instead of blue for annotated (7).}


\textit{\textbf{Theme Image Generation.}} This module goes beyond mere aesthetics and strives to bridge textual theme narratives with consistent visual image. 
It's not just about generating visuals but ensuring they are in perfect harmony with the overall design theme. 
The spike in the FID score to 33.08, when this module is excluded, accentuates its paramount significance. 
When this module is excluded, the FID score soars to 33.08, emphasizing its important meaning.
Beyond the numbers, this increase implies that while text description lays the groundwork, it's the visual representations that bring them to reality, enriching the design with visual context.
Thus, Theme Image Generation, just after Knowledge Retrieval, stands as a cornerstone in influencing the authenticity and relevance of the generated designs.

\revision{Fig.~\ref{fig:rq2-ablation} illustrates the effects of removing \emph{Theme Image Generation}. We can observe that the global theme color remains consistent because the textual description still works on the palette. However, local accessibility problems emerge. For example, in yellow boxes 4 to 5, a white icon rendered on a white search bar background becomes invisible, and other elements lack the contrast needed for a further improvement. These results demonstrate that textual cues alone are insufficient: the image module is an important module for achieving those cues in a consistent, accessible, and visually compelling interface.}

\textit{\textbf{Sub-module Execution.}} This phase ensures that suitable sub-modules are selected for collaboration, optimizing each design component's realization. Based on the evaluation of the Rico dataset, these sub-modules have specialized roles: the Text Module managed ``Text Button'' and "Text" components; the Image Module was responsible for "Image" and "Background Image"; while the Icon Module was tailored for "Icon" components. The slight variations in FID scores — 24.06(+0.3), 24.71(+0.95), and 24.32(+0.56) — when these modules were individually excluded, it became evident from their collective contributions that our fine-grained generation approach is key to driving the model's peak performance.

\revision{Ablating the \emph{Text}, \emph{Image}, or \emph{Icon} Content Modules confirms their specialized roles. Compared with full \tool{} output, green boxes 1–24 in Fig.~\ref{fig:rq2-ablation} mark missing text, image, or icon components when the corresponding module is disabled. }

\begin{myquote}
\textbf{Answer to RQ2:}
Every module in \tool{} plays a crucial role in ensuring superior design quality and diversity. The ``Knowledge Retrieval'' and ``Theme Image Generation'' modules are the most influential modules. Meanwhile, the ``Theme Description Generation'' and various ``Sub-modules'' work in concert to fine-tune and enrich the final output, adding layers of complexity and refinement to the overall design. This experiment confirms the effectiveness of our decoupling strategy, where specialized modules handle distinct components of the process.
\end{myquote}

\section{User Study}


To further evaluate the perceived usefulness (i.e., performance and usage) of our system, we carried out two user studies. The \textbf{performance} study focuses on the generated UI design compared to image generation models, and another \textbf{usage} study assesses the usage of the system compared to the latest industrial tools. 
We aim to answer the following research questions: 
\begin{itemize}
    \item \textbf{RQ3 (Performance)}: \bluetext{What is the perceived satisfaction with \tool{} compared to existing image generation tools? }
    \item \textbf{RQ4 (Usage)}:
    \bluetext{What is the perceived usefulness of our \tool{} compared to the state-of-the-art industrial tools in terms of the five findings identified in Section~\ref{sec:preliminaryStudy}?} 
    \item \revision{RQ5: (Case): What are the strengths and weaknesses of different system variants, and how can they inform future design tools?}
\end{itemize}

\textbf{Participants:}
For both user studies, We recruited the participants through alumni network and partnerships. 
As a result, we obtained 16 UI/UX practitioners (6 males and 10 females) from various corporate entities to attend our user study. 
The working experience of these professionals varies, allowing for a broad insight into industry practices and expectations. We had 8 participants with 1-3 years of experience, 5 with 3-5 years, and a further 3 who have been active in the field for more than 5 years. These practitioners come from a spectrum of roles within the UI/UX domain, including UI/UX Designers, UX Engineers, Interaction Designers, and UX Researchers.

\subsection{User Study 1: Performance Study}
\label{sec:userstudy1}

\bluetext{To assess designer satisfaction with the quality of our generated results compared to the baseline model, we conducted a performance evaluation focusing on three key aspects: functional semantics, design aesthetics, and color harmony. These criteria were selected to measure the effectiveness and appeal of the generated designs, ensuring both functional accuracy and visual coherence.}

\subsubsection{Procedure}

\bluetext{
Building on established GUI and image evaluation methods~\cite{reinecke2013predicting, zhang2008image}, participants assessed the quality of mobile GUI designs based on three metrics: functional semantics, design aesthetics, and color harmony. 
Functional semantics evaluates the relevance and clarity of the generated content, focusing on how well it aligns with the intended functions and the quality of its meaning.
Design aesthetics assessed visual appeal~\cite{zhao2021guigan}, and color harmony examined the effectiveness of color combinations. 
Each design was rated on a 5-point Likert scale. For practical evaluation, 20 design tasks from the Screen2Word dataset~\cite{2021Screen2Words} were selected, and three prototypes were generated per task using both \tool{} and stable diffusion models for comparison.
}
Participants were briefed on these metrics before evaluating the generated GUI designs. They independently scored each design, with the source model of each design concealed to maintain objectivity.

\begin{figure*}
  \centering
  \includegraphics[width=0.90\linewidth]{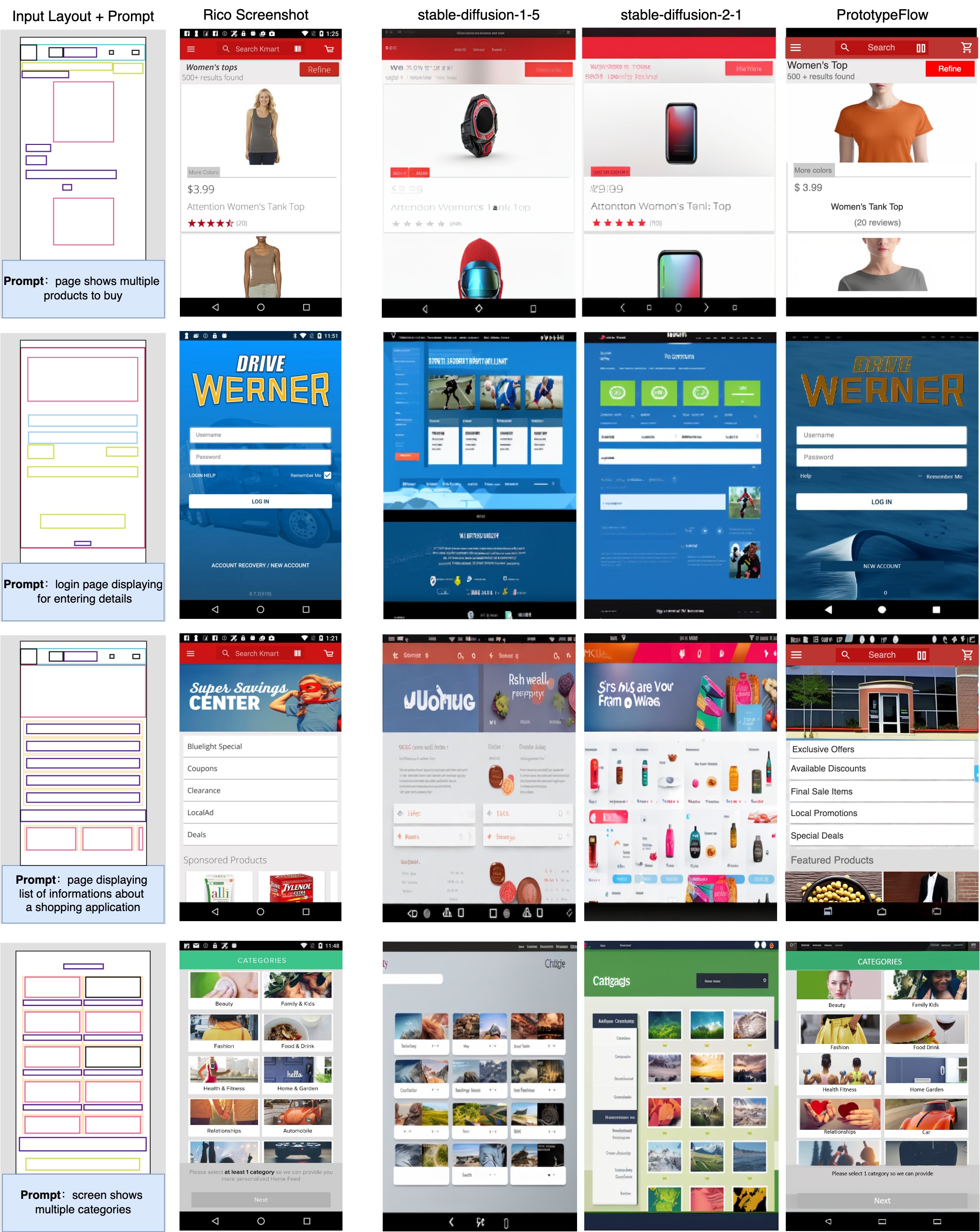}
  \caption{Generated GUI Examples from Input Layout and Prompt: Corresponding Rico Screenshots, stable-diffusion-1-5, stable-diffusion-2-1, and Our \tool{}.
  }
  \label{fig:rq2}
\end{figure*}

\subsubsection{Results \& Discussion}

\begin{figure}[t]
  \centering
  \includegraphics[width=0.7\linewidth]{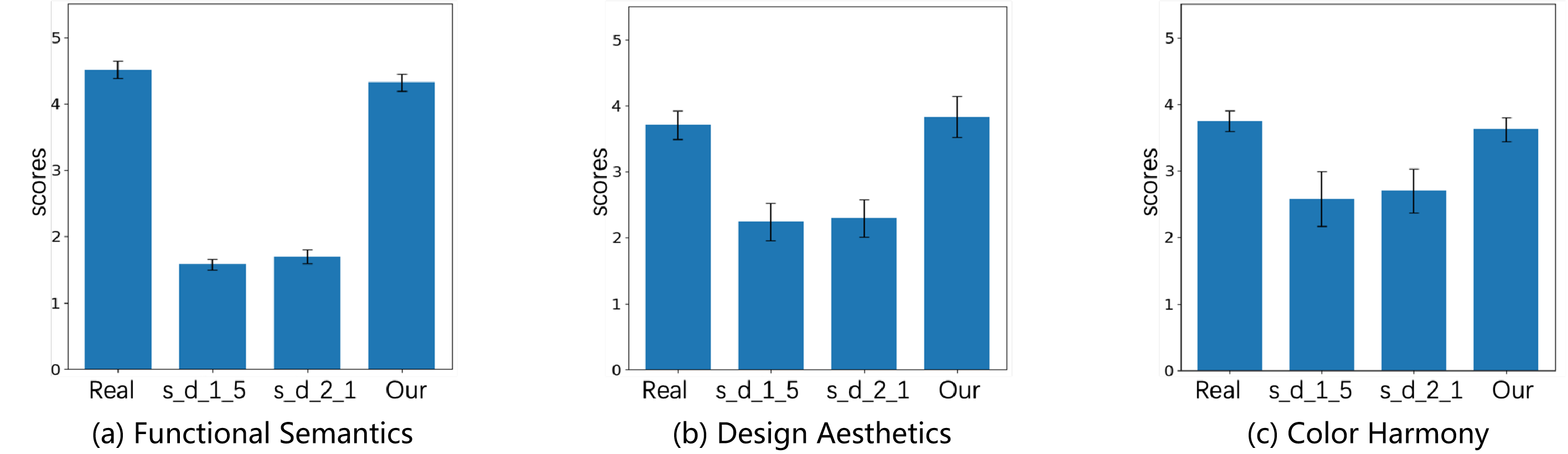}
  \caption{Comparative Evaluation of Prototype Generation. ``Real'' refers to results obtained from corresponding real screenshot images, ``s-d-1-5'' represents the outcomes of the ``stable-diffusion-1-5'' model, ``s-d-2-1'' represents the outcomes of the ``stable-diffusion-2-1'' model, and ``Our'' indicates results generated by our \tool{}. 
  }
\label{fig:userstudy}
\end{figure}

As shown in Fig.~\ref{fig:userstudy}, the GUI designs generated by our model outperformed those generated by other methods, achieving significantly higher scores in terms of design-prototype consistency (Mean=4.13), design aesthetics (Mean=3.83) and colour harmony (Mean=3.62).

\textit{\textbf{Analysis of Functional Semantics.}}
Through detailed analysis of the experimental results, we identified common characteristics of low-scoring prototype designs in terms of functional semantics, such as incomplete structures, basic content, overly small or abrupt images, and an excessive number of components. 
In contrast, high-scoring prototype designs had a clean layout, moderately rich content, and compatible images. The balance between content richness and layout simplicity was also highlighted as an important consideration.

\bluetext{Furthermore, we found that most of our results demonstrated accurate semantic alignment with the design intent, due to our decoupled generation mechanism.} This was evident from the high score of 4.32. Comparatively, this score was significantly higher than that of the stable-diffusion-1-5 and stable-diffusion-2-1 generated results 1.58 and 1.7, yet just 0.19 score lower than the real screenshots (4.51). As demonstrated in Fig.~\ref{fig:rq2}, the stable-diffusion models largely presented images lacking specific semantic content, thus appearing blurry. 

\textit{\textbf{Analysis of Design Aesthetics and Color Harmony.}}
Interestingly, a few of the prototypes generated by our model scored higher than real-world prototypes - a notable accomplishment when compared against the benchmark of real screenshots. 
In Fig.~\ref{fig:userstudy} (b), our model surpassed the average score for real prototypes in terms of design aesthetics (3.83 compared to 3.71) and in Fig.~\ref{fig:userstudy} (c) was slightly lower with respect to colour harmony (3.62 compared to 3.74). This is a significant improvement over the stable-diffusion models. Our generated prototypes closely mirrored real-world prototypes' overall aesthetics and colour harmony and, in some instances, were of superior quality to poorly designed real-world prototypes. 

\textit{\textbf{Analysis of Failure Case.}}
Upon cross-verification of our model's results against real images, we found some minor errors of our results, such as the over-generation of text leading to typography issues. Participant feedback suggested the further alignment of our model with typography and other design guidelines could address this issue, enhancing the generation performance. 

\begin{myquote}
\textbf{Answer to RQ3:}
The user study underscored the efficacy of \tool{} in creating GUI designs that excel in design-prototype consistency and aesthetic appeal and colour harmony compared to the state-of-the-art image generation models. Notably, the outputs were found to align closely with real-world screenshots, even outperforming them in terms of design aesthetics. Further analysis underscored that our \tool{} decoupled generation mechanism achieved accurate semantic alignment with the design intent.
\end{myquote}

\subsection{{User Study 2: Usage Study}}

\revision{In order to obtain feedback and facilitate further discussion on our findings and tool usage, we compared our \tool{} with emerging design tools; we conducted another user study followed by expert interviews. This study involved a comparative analysis using our tool, Vercel's V0\cite{v0dev2024}—a UI code-based design tool, and Uizard\cite{uizard}—a Prompt to UI design tool.} After engaging with these tools, participants completed a 5-point Likert Scale questionnaire based on 5 questions. The questionnaire aimed to assess how well our system addressed five design goals identified in Section \ref{sec:preliminaryStudy}, in terms of all aspects of performance and interactiveness.

\subsubsection{Procedure}

The study began with an introductory session to acquaint participants with the study procedure. Participants were then asked to create 2 GUI prototypes using our tool, Vercel's V0, and Uizard, based on provided design purposes and corresponding wireframes. Those 2 prototype design purposes are random obtained from 20 different design purposes, which are selected from the Screen2Word test dataset for GUI generation, with each containing an average of 5 components.
\bluetext{Participants were allowed to modify the input description until they were satisfied with the generated results, and the number of manual modifications made by users was recorded.}

Following task completion, we conducted a semi-structured survey with each participant. This survey included 5 main questions on a 5-point Likert Scale, focusing on eliciting feedback about our tool's performance relative to the five identified design findings. Additionally, participants were asked for suggestions on potential improvements. The survey questions were as follows:

\begin{enumerate}
    \item \revision{How effective was the system in providing relevant design suggestions or resources from the knowledge base?
    (Related to F1: Knowledge Base Effectiveness)}
    \item \revision{Was the input method intuitive, and did it allow you to express the design intent clearly and precisely?
    (Related to F2: More Input Control and Flexible Output Editability in Design Generation Process)}
    \item \revision{Did the tool provide useful suggestions to clarify your design intent?
    (Related to F3: Supporting Designers in Expressing Intent Through Prompts)}
    \item \revision{Were the intermediate results useful in aligning the design with your intent, and did they enhance interaction?
    (Related to F4: Precise Control in Generation Processes)}
    \item \revision{Did the tool maintain consistent semantics and style across all design elements, ensuring a cohesive output?
    (Related to F5:  Maintaining Thematic Consistency and Coherence Across Generated Components)}
\end{enumerate}

\begin{figure}
    \centering
    \includegraphics[width=0.9\linewidth]{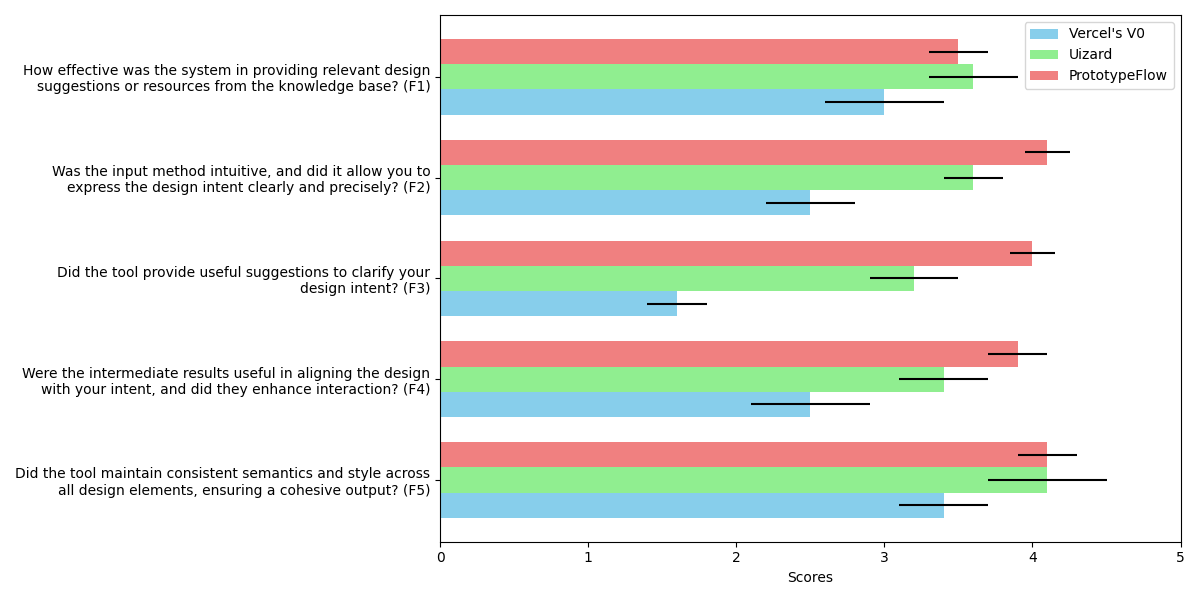}
    \caption{Results of the usage study.
    Blue represents Vercel's V0, green denotes Uizard, and red indicates our system \tool. 
    }
    \vspace{-2mm}
    \label{fig:u2result}
\end{figure}
\vspace{-2mm}

\subsubsection{Results \& Discussion}
The feedback gathered from our semi-structured survey offered critical insights into the performance of our tool in relation to the challenges outlined in Section~\ref{sec:preliminaryStudy}, and highlighted potential areas for enhancement. 
Fig. \ref{fig:u2result} shows the results. 

\revision{For statistical analysis, we conducted independent-samples t-tests comparing \tool{} with Vercel’s V0 and Uizard across five user survey question criteria (F1–F5). Results show that \tool{} significantly outperformed Vercel on all dimensions (p < 0.001). Compared to Uizard, \tool{} showed significant advantages in intermediate transparent and editable generation (F4), suggestion usefulness in clarifying and enhancing designer prompts (F3), and input expressiveness (F2), all with p < 0.001. For design theme consistency (F5), \tool{} and Uizard performed similarly (p = 0.071), and for relevance of retrieved suggestions (F1), \tool{} scored slightly lower (3.5 vs. 3.6, p = 0.4). These findings indicate that \tool{} excels at enhancing clarity, interaction, and refinement in the design process, while maintaining comparable output consistency.}

\textit{\textbf{Knowledge Base Effectiveness (F1).}} Participants rated the Relevant Designs, with Uizard achieving a score of 3.5, our tool closely following at 3.4, and Vercel's V0 receiving 3.0. 
\bluetext{In follow-up interviews, we found that 9/16 participants believed our tool excelled in generating detailed elements, such as text, icons, and images. This feedback highlights the strength of our decoupled generation method, which maximizes the system’s ability to create detailed, functional design elements.}
\bluetext{On the other hand, 7/16 participants appreciated Uizard’s overall style design, describing it as visually more appealing. }
\bluetext{In addition, 4/16 participants praised our system's flexibility, particularly the ability to import and integrate their own data into the generation process, allowing for customization. They suggested that the effectiveness of this feature could be verified in the future.}

\bluetext{This feedback highlights the potential of our tool in integrating design libraries—a feature lacking in Vercel's V0 and partially implemented in Uizard. Our \tool{} offers a more customizable and functional design experience.}

\textit{\textbf{\revision{More Input Control and Flexible Output Editability in Design Generation Process (F2).}}} Our system led in reflecting design intent with a score of 4.1, compared to Uizard's 3.6 and Vercel’s 2.5. Participants valued our tool's use of both wireframes and prompts, which Uizard lacks, and found Vercel’s code generation to not adhere closely to layout sizes. A participant remarked, \textit{``\tool{}'s method of rendering components from wireframes and prompts is impressive. The design of layout parts encompasses functional design and is labor-intensive. AI assistance in this area is much needed.''} 

\textit{\textbf{\revision{Supporting Designers in Expressing Intent Through Prompts (F3).}}}
\bluetext{
Our \tool{} scored 4.0 for Prompt Enhancement, outperforming Uizard (3.2) and Vercel's V0 (1.6). 14/16 participants found our clarification ability particularly helpful, especially for themes and app categories. One participant noted, "The automatic clarification is more professional than my own input, saving me from guessing what the AI needs." This feature allowed designers to focus more on achieving their desired results, reducing guesswork and effort.
On average, designers using our tool made 2.6 revisions to reach their desired design. In comparison, Uizard required 4.3 revisions, and Vercel's V0 required 6.1 revisions, with many users still unsatisfied, citing lack of control over the generated code format.}

\bluetext{
Regarding color clarification, designers with more than three years of experience expressed the need for tools that respect company design constraints, particularly for theme colors. They suggested that future improvements could include interactive visualizations with customizable color palettes to enable more precise adjustments.
}

\textit{\textbf{\revision{Precise Control in Generation Processes (F4).}}}
\revision{Our tool was highly regarded for its intermediate generation process in helping with human-in-the-loop interaction, achieving a score of 3.8. }In comparison, Uizard scored 3.5, while Vercel's V0 lagged behind with a 2.5. Notably, 14 out of 16 participants found the intermediate steps in our tool exceedingly beneficial for interactive engagement with the AI. A participant shared a compelling example: \textit{``While working on a job search page, I was impressed by the UI designs produced by the tool. They were professional and included all necessary details, such as job categories, locations, and salaries, making it feel like a real, complete design.''} This feedback underscores our tool's proficiency in providing clear, informative intermediate steps that facilitate effective user interaction and ensure the generation of high-quality prototypes.

Participants highlighted a distinctive advantage of our tool over others: while Vercel's V0 allows code-based component edits and Uizard enables manual adjustments, our tool uniquely presents both the components and critical information during their generation, including prompts. This feature was particularly lauded for its precision and convenience. One participant expressed, \textit{``I was surprised and pleased by the direct control over fine-tuning local components through prompts. This is a stark contrast to Vercel's V0, which often struggles with accurately locating specific components for adjustments.''} Echoing this sentiment, 15 out of 16 participants appreciated our tool’s top-down editing support, which they found met their needs from broader adjustments to more detailed refinements, significantly aiding them in realizing their design vision accurately.

\textit{\textbf{\revision{ Maintaining Thematic Consistency and Coherence Across Generated Components (F5).}} }
In the critical aspect of maintaining design coherence, our tool and Uizard both achieved a high score of 4.2, clearly outperforming Vercel's V0’s 3.5. A majority of the participants, 10 out of 16, remarked on the effective maintenance of colour and image style consistency by both our tool and Uizard. They noted that these tools produced designs with cohesive colour schemes and consistent visual elements. 

However, while Vercel V0's outputs were generally competent, its lack of image components and tendency towards simplistic black and white colour schemes were seen as areas needing improvement. Beyond the aesthetic aspects of design consistency, participants also highlighted the importance of colour accessibility. A participant pointed out a significant oversight in current tools, stating, \textit{``Although the tools offer impressive theme colour designs, a key shortfall is their lack of consideration for colour accessibility, making some designs impractical for real-world application.''} This feedback underscores the necessity for future tools to incorporate colour accessibility as a fundamental component of design consistency.

\begin{myquote}
\textbf{Answer to RQ4:}
The usage study underscored the effectiveness of our tool in refining the UI prototyping process. It adeptly balances automation with customization and offers user-friendly interactions. Particularly notable is its capability to better communicate design intent and provide transparent, explainable AI-assisted intermediate results. This presentation of both components and vital information during generation has been highly praised for its precision and convenience, facilitating designer workflows more effectively than other emerging design tools.
\end{myquote}

\subsubsection{\textbf{Weaknesses and improvements.} }
Participants offered valuable feedback on potential improvements and identified exciting directions for future enhancements. Several participants saw great potential in augmenting the user experience by introducing automated wireframe generation from high-level descriptions. This feature could significantly streamline the design process by reducing manual steps, thereby increasing efficiency. A participant highlighted this by saying, \textit{``It would be really helpful if the tool could auto-generate wireframes based on high-level descriptions. It would save me more time.''}

Addressing specific failure cases, one participant pointed out a drawback in the current image generation process: \textit{``Sometimes the generated images for small, simple areas are overly complex. Simplified vector images like icons might be more effective.''} This observation led to the suggestion that the tool should differentiate between standard images and vector graphics in future iterations for improved functionality.

Furthermore, in terms of usability and accessibility, participants suggested that aligning the tool more closely with established UI design guidelines, such as Material Design, would greatly enhance its utility. One participant emphasized the importance of functionality alongside aesthetics: \textit{``Enhancing the tool to adhere to UI design guidelines, such as Material Design, would significantly improve its usability. A prototype should not only look good but should also be interactive and user-friendly.''} Building upon this, another added, \textit{``Creating visually stunning interfaces is great, but incorporating accessibility guidelines is crucial for making designs truly universal.''}

\subsection{\revision{User Study 3: Case Study}}
\label{sec:casestudy}

\revision{To explore greater flexibility and understand the trade-off between precise control and design flexibility, we introduce a case study that envisions idealized variations of our system and examines their strengths, limitations, and potential directions for future development.}

\subsubsection{\textbf{\revision{Experimental Setup.}}} \revision{We explore four types of variations across both input and output dimensions. 
For \textbf{input variants}, we consider \textbf{(1) Input-NoLayout} – removing the layout constraint to allow generation of UIs that preserve semantic structure without adhering to a fixed layout; \textbf{(2) Input-NoKnowledge} – removing the knowledge reference to enable the generation of more diverse and unconstrained applications. 
For \textbf{output variants}, we consider \textbf{(3) Output-MultiLayout} – generating multiple UI prototypes with varying layouts for the same semantic and thematic specification; \textbf{(4) Output-MultiTheme} – generating UI prototypes that maintain consistent layout and semantics but vary in thematic styles. }

\begin{figure*}
  \centering
  \includegraphics[width=1\linewidth]{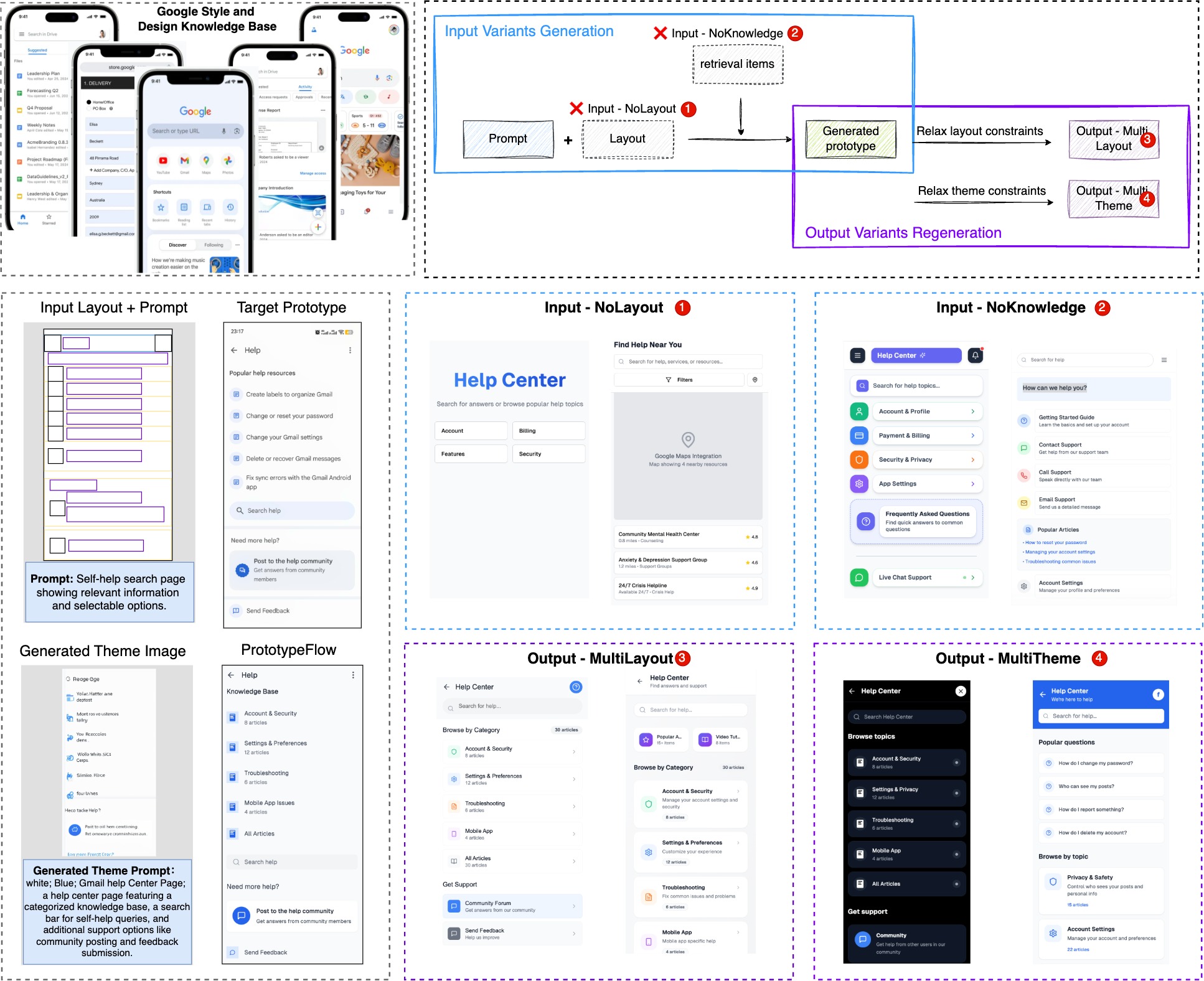}
  \caption{Illustration of ideal outputs from different variants of our \tool{}.
  }
  \label{fig:discussion}
\end{figure*}

\revision{To ground these variations, we apply our system to a scenario of designing a self-help search page for a Google product. For each variant, we construct a representative and idealized output that reflects the relaxed constraint as shown in Fig.~\ref{fig:discussion}. These outputs serve as design probes to support structured reflection. We then re-engaged the same 16 designers from our previous study and invited them to evaluate each variation and compare. They were asked to reflect on the perceived value, applicable design stage, and trade-offs of each alternative.
After obtaining feedback from the designers, we conducted a thematic analysis of the interview data. All responses were transcribed and open-coded to identify patterns in designers’ perceptions of each variant’s usefulness, strengths, and limitations. Two researchers independently coded a subset of the data and resolved differences through discussion to ensure consistency.}

\subsubsection{\textbf{\revision{Findings.}}} 
\revision{After analyzing the feedback and the open-coding methods, we derived 3 key findings:}

\revision{\textbf{Input Preferences During Ideation:} 13 out of 15 designers—especially those with over three years of experience—expressed a preference for the Input–NoKnowledge variant during the early ideation stage. They felt that removing the knowledge constraint allowed for broader and more diverse design suggestions, which helped stimulate creative exploration. These designers noted that relying solely on the internal knowledge base at this stage tended to limit novelty and inspiration. By contrast, the Input–NoLayout variant was perceived as less beneficial during ideation, as removing layout constraints often resulted in outputs that were too detached from practical or implementable UI structures.}

\revision{\textbf{Wireframes as a Preferred Input Modality:} Designers consistently favored wireframes as their primary method of interaction, emphasizing that this input format aligns with their established workflows and offers precise control over layout and structure. Many described wireframes as a “native language” for communicating design intent—more intuitive and efficient than crafting detailed textual prompts. Several participants noted they maintained personal libraries of reusable wireframes, enabling rapid adaptation to new design tasks. This strong preference reflects a desire to retain detailed layout control by default, and only strategically apply variants like Input–NoLayout when seeking greater output diversity.}

\revision{\textbf{Output Diversity Preferences: Helpful for Learning, Less Aligned with Team Workflows.} The Output–MultiLayout and Output–MultiTheme variants—representing post-generation diversity—were appreciated primarily by less-experienced designers (4 out of 15, with fewer than two years of experience). For these participants, generating multiple design alternatives after the initial prototype served as a valuable learning tool, enabling comparison, self-validation, and a deeper understanding of design trade-offs. However, even these designers noted that such diversity is less applicable in real-world team settings, where key layout and functionality decisions are typically made collaboratively before generation begins. Most experienced designers (11 out of 15) shared this view, emphasizing that once a team reaches alignment, generating additional alternatives can distract from execution and reduce efficiency. They preferred to front-load diversity during early ideation (e.g., through Input–NoKnowledge), and viewed post-generation variation as potentially disruptive to streamlined workflows and delivery timelines.}

\begin{myquote}
\textbf{Answer to RQ5:}
\revision{Our findings indicate that the timing and form of diversity introduction should align with the designer’s experience and workflow context. Input-level variants like Input–NoKnowledge were preferred by experienced designers during early ideation, enabling broader exploration. In contrast, output-level variants such as Output–MultiLayout and Output–MultiTheme were more helpful for novice designers to compare alternatives and learn from variation. However, these output variants were seen as less useful in team-based settings, where design decisions are often finalized early. These results suggest that future systems should allow dynamic adjustment between control and flexibility, tailored to users’ experience levels and design stages.}
\end{myquote}

\section{Discussion and Future work}

\revision{The rapid rise of generative models and prompt engineering is transforming many landscapes, including design tools. Our findings highlight the need for GenAI design systems that balance fine-grained designer control with diverse output and adapt to a broad spectrum of users and design goals. Managing and updating design knowledge bases remains a challenge, as static resources can quickly become outdated in fast-evolving contexts. This section discusses key implications from our system design and user studies and identifies future directions, including  broader user and platform support, dynamic knowledge management, and enhanced adaptability to emerging design trends.}

\revision{\subsection{Tailoring Interaction Modalities to User Expertise}}

\revision{Our research primarily targeted professional UI/UX designers and found that these experts prefer to interact with AI design tools in their own “native” design languages, primarily wireframes and visual layouts. They noted that this approach gives them precise control over structure and aligns with their established workflows. One interviewee described wireframes as the ``native language'' for communicating design intent, more intuitive than lengthy text prompts. This insight underscores a key design principle: AI tools should ``align with designers’ visual habits'' and not force them into unfamiliar interaction styles\cite{khan2025beyond}. In practice, this means integrating GenAI assistance into existing GUI design environments (e.g. Figma, Adobe XD) where designers naturally work with visual components, rather than requiring designers to write long textual descriptions of UIs. Indeed, recent studies have found that designers want AI helpers to ``go beyond textual prompting'' and seamlessly fit into visual ideation processes\cite{khan2025beyond}. Our \tool{} reflects this by letting designers sketch a layout or provide a wireframe which the system then fleshes out, rather than relying on abstract prompts alone.}

\revision{However, not all users of AI-assisted design tools will be professionals. Novice designers or non-designers, such as developers or small entrepreneurs, may lack the training to sketch wireframes or use design terminology. For these users, the system must ``speak'' a different language – one that matches their mental model and skills. Prior work has identified this gap: only focusing on experienced designers leaves out the perspectives of novice designers and students, whose needs and ways of communicating with AI might differ significantly \cite{khan2025beyond}. For example, a beginner might prefer describing the intended interface in plain language (e.g. ``I need an e-commerce home page with a search bar at the top and product cards'') or selecting from example images or templates, rather than drawing a layout from scratch, as they don't have their own wireframe design library. There is potential for GenAI design tools to incorporate multiple input modalities to accommodate this range of users. A system could allow a user to start from a high-level natural-language description or a storyboard of screenshots, and then iteratively refine the design with more visual inputs as their confidence grows. In essence, the AI should serve as a fluent translator between different ``UI languages'' – whether that’s the precise vocabulary of a UX professional or the rough descriptions a novice might give. By tailoring the interaction style (textual, visual, conversational, or guidance-based) to the user’s expertise, we can lower the barrier to entry and democratize the design process. Our findings that experts prefer wireframe-driven interaction suggests one size won’t fit all; future systems might include an adaptive interface that, for instance, starts with a friendly Q\&A or template gallery for novices and gradually introduces more free-form wireframing as the user gains proficiency. Such adaptability would help non-professionals work in whatever way feels most natural to them, turning AI into a true creative partner rather than a rigid tool.}

\revision{\subsection{Balancing Fine-Grained Control and Output Diversity in Generative Design}}

\revision{A recurring topic is the tension between giving designers fine-grained control over outputs, while providing high output diversity for inspiration. Our findings reflect this balance: senior designers tend to prioritize control and precision, especially in later stages of design (such as creation and iteration), whereas less-experienced designers or early ideation phases benefit from a breadth of diverse suggestions. In our case study, 13 of 15 designers preferred turning off the internal knowledge base during early ideation to avoid constraining the AI to known patterns. This allowed more divergent and novel UI suggestions to emerge. By contrast, turning off layout constraints was seen as less useful at creation – outputs became too unimplementable. This highlights that designers desire structured diversity: room for exploration, but within workable bounds. }

\revision{Prior work echoes this need for multiple options and iterative exploration; for example, novice spatial designers wished for AI systems to generate multiple design options, so they can choose and refine among them\cite{wan2024breaking}. At the same time, users criticized fully automated, ``one-shot'' generation as a black box process that they couldn’t intervene in, preferring approaches (like retrieval-based methods) that allowed more control over intermediate steps\cite{wan2024breaking}. In the context of UI design, this means an effective GenAI tool should support both divergent exploration and convergent refinement. This aligns with the classic divergent–convergent model of design thinking, the system should encourage divergence (high variety, novelty), and later enable convergence (fine-tuning a chosen direction). Recent HCI research emphasizes balancing these dual needs: designers call for AI tools that help ``balance efficiency and exploration'', providing varied design alternatives while still letting them control the process. }

\revision{Importantly, the value of output diversity appears context-dependent. Our participants noted that generating many alternative layouts or themes after an initial prototype was mainly useful as a learning aid for junior designers (e.g. to compare options and understand trade-offs), but was less aligned with the realities of professional team workflows. Once a design direction is decided collaboratively, excessive variations can become a distraction and slow down execution. Experienced designers in our study preferred to front-load diversity in the early ideation stage to broaden horizons, then commit to a direction and focus on polishing it. This sentiment resonates with findings by Khan et al.\cite{khan2025beyond}: professional UI/UX designers ``valued AI tools that offer greater control over ideation'' while still generating design alternatives for inspiration. In sum, an AI design system must intelligently support both modes – offering divergent idea generation when appropriate and convergent fine-tuning when the designer needs precision. Designing interfaces that let users fluidly toggle or transition between these modes (for example, a ``surprise me with alternatives'' button versus a ``lock this layout'' mode) could be a promising direction for future tools.}

\revision{\subsection{The Limits of Static Knowledge Bases in Rapidly Evolving Design Contexts}}

\revision{Our \tool{} system features a knowledge-based retrieval module that leverages a local repository of established design patterns and examples (such as company style guidelines) to ground generative outputs in familiar, brand-consistent solutions. This foundation helps ensure results are realistic and on-brand, accelerating routine design tasks with proven patterns. However, our interviews highlight several limitations to this approach, particularly as design contexts and trends rapidly evolve. First, reliance on a static or narrowly scoped knowledge base can quickly lead to outdated or repetitive suggestions, constraining creativity during early ideation. Designers noted in Section~\ref{sec:casestudy} that while knowledge-based retrieval is helpful for efficiency, over-dependence early in the process risks limiting exploration and novelty. This points to an inherent trade-off between exploiting known best practices and enabling creative divergence. A local or company-specific knowledge base that is too limited or poorly maintained may cause the system to repeatedly generate similar styles, ultimately missing out on new design paradigms and innovative alternatives.}

\revision{Second, maintaining the relevance and quality of the knowledge base is an ongoing challenge. In real-world team knowledge base deployments, questions arise: Who is responsible for updating the design repository? How are new styles and patterns integrated? If left unattended, the system may propose outdated trends (such as the overuse of gradients or drop shadows), reducing the value of its recommendations. Furthermore, adaptability to diverse design contexts remains an open issue. A knowledge base primarily built from e-commerce app screens may not generalize well to other domains, such as data visualization dashboards or virtual reality interfaces. Flexible retrieval that allows users to constrain or expand the knowledge base according to project needs may help address this, and our system takes a first step by enabling users to toggle knowledge use depending on the task. However, more automation and standardization mechanisms for dynamic updating and domain adaptation are needed.}

\revision{To address these limitations, future work should explore more dynamic and scalable approaches to knowledge management. For example, integrating external sources such as public design repositories, design galleries, or open-source community platforms (similar to Hugging Face in deep learning) could help ensure broader coverage and more timely updates. Establishing standardized formats for online management and community contributions would further support the evolution and sharing of up-to-date design knowledge.}

\revision{In summary, while knowledge-based UI generation provides a strong foundation and alignment with established design standards, its long-term effectiveness depends on broad coverage, timely updates, adaptability to new domains, and the ability to learn from ongoing user feedback. Ensuring that GenAI design assistants stay in sync with current trends and remain responsive to diverse project contexts will be key to their sustained usefulness.}

\revision{\subsection{Integrating GUI and Web Design Practices for Broader Design Knowledge}}

\revision{While PrototypeFlow and similar generative AI systems have advanced graphical user interface (GUI) design, particularly for mobile and desktop app screens, it is important to consider how these approaches might extend to web design and other platforms. Designing for the web introduces unique challenges and requires a different mindset from traditional software UI design. As Jakob Nielsen observed\cite{nielsen1997difference}, web designers cannot exercise full control over the user interface, since the end-user’s device, browser, and personal settings all influence the final appearance and behavior. In practice, this means web layouts must be fluid and adaptable to various screen sizes, network conditions, and accessibility requirements, whereas GUI designers for native applications typically operate with more predictable parameters and can specify layouts with greater precision.This fundamental difference highlights a key limitation of current GenAI design tools trained predominantly on mobile app screens or static interfaces. Such systems may not generalize well to web design scenarios, which demand features like responsive layout reflow, hyperlink navigation, and compliance with web standards such as HTML, CSS, and accessibility guidelines. By contrast, GUI design tools often rely on platform-specific conventions, like those prescribed in the iOS Human Interface Guidelines or Android Material Design.}

\revision{To bridge these domains, a GenAI design system should incorporate context awareness, enabling it to recognize the intended medium and apply design patterns suited to that context. For example, an AI tool generating a website should suggest responsive grid systems and ensure that layouts adapt seamlessly from mobile to desktop. In contrast, for mobile app design, the system might prioritize native components and platform-specific navigation. Prior research in HCI has shown the value of such context sensitivity. For instance, Landay’s DENIM tool~\cite{lin2001denim} was developed specifically for early-stage web design, integrating site maps and page flows to address web designers’ needs—needs that were not met by conventional GUI prototyping tools. This example underscores that certain aspects of design knowledge are inherently tied to the target domain.}

\revision{Despite these differences, some principles—such as visual hierarchy, consistency, and affordances—are universal across both GUI and web design. The next generation of AI design assistants could leverage a broad foundation of design knowledge while dynamically adapting recommendations and outputs to suit the relevant context. Currently, our system focuses on GUI prototyping. To support web design, it would need to accommodate multi-page navigation, fluid layouts, and a wider set of building blocks such as form elements and navigation bars. Achieving true cross-platform capability will require training on diverse datasets, including both mobile app UIs and responsive web designs, as well as offering user options to tailor the generation process to specific platform requirements. By acknowledging and embracing these domain differences, and drawing from expert guidance on design adaptability, future GenAI tools can become more general partners for designers, supporting creative work across an expanding range of digital interfaces.}

\section{Conclusion}
In this paper, we identified five key gaps in designers’ workflows with current AI-assisted design tools. To address these challenges, we introduced \tool{}, a novel multi-module system that balances automation with customization. 
Given human-provided descriptions and wireframe layouts, our system iteratively refines these inputs into engaging, high-fidelity design prototypes, maintaining aesthetic harmony and aligning with the design intent. Beyond generation,
\revision{\tool{} not only automates design intent enhancement for designers but also provides editable intermediate results to enhance rapid regeneration.}
Our quantitative and qualitative evaluations further corroborate the potential of our approach to significantly improve the UI / UX design process. 
Going forward, we will continue improving our work, like enabling automated wireframe generation and supporting dynamic component integration for a more universal, user-friendly, efficient and creative design process.

\bibliographystyle{ACM-Reference-Format}
\bibliography{reference}

\end{document}